# A general-purpose neural network potential for Ti-Al-Nb alloys towards large-scale molecular dynamics with *ab initio* accuracy

Zhiqiang Zhao, Wanlin Guo[*] and Zhuhua Zhang[*]

*State Key Laboratory of Mechanics and Control for Aerospace Structures, Key Laboratory for Intelligent Nano Materials and Devices of Ministry of Education, and Institute for Frontier Science, Nanjing University of Aeronautics and Astronautics, Nanjing 210016, China*

Email: wlguo@nuaa.edu.cn; chuwazhang@nuaa.edu.cn

**Abstract**: High Nb-containing TiAl alloys exhibit exceptional high-temperature strength and room-temperature ductility, making them widely used in hot-section components of automotive and aerospace engines. However, the lack of accurate interatomic interaction potentials for large-scale modeling severely hampers a comprehensive understanding of the failure mechanism of Ti-Al-Nb alloys and the development of strategies to enhance the mechanical properties. Here, we develop a general-purpose machine-learned potential (MLP) for the Ti-Al-Nb ternary system by combining the neural evolution potentials framework with an active learning scheme. The developed MLP, trained on extensive first-principles datasets, demonstrates remarkable accuracy in predicting various lattice and defect properties, as well as high-temperature characteristics such as thermal expansion and melting point for TiAl systems. Notably, this potential can effectively describe the key effect of Nb doping on stacking fault energies and formation energies. Of practical importance is that our MLP enables large-scale molecular dynamics simulations involving tens of millions of atoms with *ab initio* accuracy, achieving an outstanding balance between computational speed and accuracy. These results pave the way for studying micro-mechanical behaviors in TiAl lamellar structures and developing high-performance TiAl alloys towards applications at elevated temperatures.

# I. INTRODUCTION

TiAl intermetallic compounds possess a number of desirable properties, including low density (3.9-4.2 g/cm³), high melting points, exceptional oxidation resistance, and high-temperature creep resistance [1,2], rendering them excellent candidates for developing high-temperature structural materials towards advanced applications. Notably, γ-TiAl (face-centered tetragonal, $L1_0$) and $α_2$-$Ti_3Al$ (hexagonal, $D0_19$) alloys play pivotal roles in the aerospace sector, particularly in meeting the stringent demands of gas turbine engines. For example, using TiAl-4822 to substitute the Ni-based high-temperature alloy in the low-pressure turbine blades of Boeing 787 led to a weight reduction of approximately 200 kg and significant improvements in propulsion efficiency [3,4]. However, the applications of TiAl alloys are still hampered by several drawbacks, such as low ductility at room temperature and inferior high-temperature strength [2,5]. Over more than half a century, concerted research efforts have been directed toward enhancing the mechanical properties of TiAl alloys to accommodate increasingly demanding requirements of service temperatures [6].

Recently, polysynthetic twinned (PST) TiAl single crystals with controlled lamellar orientations have been successfully fabricated by directional solidification [7]. The new PST-TiAl single crystals (Ti-45Al-8Nb) exhibit a remarkable combination of strength, ductility, and creep resistance. Compared with TiAl-4822 alloy [8], samples with 0° lamellar orientation exhibit an increased average room temperature tensile ductility from 2% to 6.9%, and an increased yield strength from less than 450 MPa to greater than 700 MPa [7]. More importantly, at 900°C, the yield strength of PST-TiAl single crystals still stays at 637 MPa with an elongation of 8.1% [7]. The introduction of Nb reduces the stacking fault energy, making the dislocations sweep more easily to form twins during deformation [7,9]. The generation of deformation twins simultaneously effectively refines the lamellae, ultimately leading to the significant improvements in ductility and strength [5,10,11]. Therefore, studying the high-temperature mechanical properties of TiAl alloys with high Nb content and investigating the influence of Nb on the strength and ductility of these alloys are of great significance.

Several experimental studies [12-17] have explored the contribution of Nb to the toughness and yield strength of TiAl-based alloys, but the effect of Nb on the high-temperature properties of these alloys remains unclear. Moreover, since the TiAl alloys have unique microstructure, experimental observations of their mechanical deformations are challenging. Computational simulations can provide crucial insights into the plasticity mechanisms of TiAl alloys. For instance, density functional theory (DFT) calculations have suggested strong preferences of Nb atoms to occupy the Ti sites [18,19] in both single phases [20] and at phase interfaces [21]. The DFT computations are limited to systems with hundreds of atoms, while incompetent for studying complex dynamic dislocation slip mechanisms of alloy systems. To overcome these limitations, many classical potentials including embedded atom method (EAM) [22,23] and modified embedded atom method (MEAM) [24-26] potentials have been developed for the Ti-Al binary [22,24,25] and Ti-Al-Nb ternary [23,26] systems.

Nonetheless, as reported by Pei *et al*.[27], these classical potentials have considerable deficiencies in reproducing the elastic constants and thermal expansion behavior in γ-TiAl and $α_2$-Ti$_3$Al. Specifically, the EAM and MEAM potential fail to reproduce the negative Cauchy pressure in γ-TiAl and provide incorrect stacking fault energy ordering [27], thereby undermining the reliability of simulations.

In recent years, machine-learned potentials (MLPs) [28-40] have emerged as a powerful tool to address the limitations of classical potentials and accurately describe mechanical behavior in diverse elemental [41-44] and multicomponent metal systems [45-47]. For the Ti-Al binary system, notable advancements include the development of the moment tensor potential model [48] and pareto optimal model [49]. In addition, Lu *et al*. also developed a Ti-Al-Nb ternary model based on the deep potential framework [50]. However, their study only validated the ground state and defect properties, leaving the finite temperature and high-temperature properties of TiAl alloys unexplored. In general, these MLPs have significantly improved accuracy in describing the physical properties of TiAl alloys. However, MLPs typically suffer from an issue that the computational speeds are several orders of magnitude slower than classical potentials like EAM, thereby impeding large-scale simulations. Recently, Fan *et al*. proposed a MLP framework named neuroevolution potential (NEP) [39,51,52], which not only achieves *ab initio* accuracy but also offers computational speed on par with those of classical EAM and MEAM potentials. The NEP model demonstrates approximately $2.4 \times 10^7$ atom step/s using a single A100 GPU for molecular dynamics (MD) simulations [44], showing an excellent balance between accuracy and speed. This makes NEP highly suitable for developing efficient and accurate interatomic interaction models for important structural materials like TiAl-based alloys.

In this work, we develop an advanced MLP for the Ti-Al-Nb ternary system by combining the NEP framework with an active learning scheme. By pre-training on an initial DFT dataset and iteratively refining the NEP model using active learning with farthest point sampling, we can systematically enhance the MLP's robustness and predictive capabilities. In the end, using thousands of training sets, a universal MLP for the Ti-Al-Nb ternary system has been successfully constructed that is powerful for describing high-temperature mechanical properties. Our NEP model exhibits superior accuracy compared to all previous classical potentials of TiAl systems. It can reliably reproduce the following properties: 1) basic lattice and defect properties such as equations of state curves, phonon dispersions, elastic constant tensors, defect formation energies, surface energies, and interface energies; 2) temperature dependence of lattice constants and elastic constants, and high-temperature properties such as melting points; 3) γ lines and γ surfaces as well as crucial stacking fault energy ordering; 4) the influence of Nb doping on formation energy and stacking fault energy of TiAl systems. This NEP model features *ab initio* accuracy, enables MD speed comparable to classical potentials, and can scale up the MD simulations to millions of atoms. Our methodology applied to Ti-Al-Nb is general and applicable to training interatomic potentials of all elements to effectively simulate the physical and mechanical properties of various materials.

In the following, we begin with a brief introduction to the NEP framework [39,51,52], followed by a description of the tunable training hyperparameters in Section II. Subsequently, we provide an overview of the training dataset generation and DFT calculation settings. In Section III, we present the workflow for fitting the NEP model with an active learning scheme. Furthermore, we systematically compare the physical properties calculated by NEP with those obtained from DFT and other available experimental data. The calculated physical properties are comprehensive, including surface energies, vacancy formation energies, elastic constants, stacking fault energies, interface energies, finite-temperature thermodynamic properties, and so on. Finally, the study is summarized in Section IV.

## II. METHODOLOGY

### A. The NEP framework

We here used the NEP [39,51,52] framework to train the ternary Ti-Al-Nb model. The NEP framework is implemented in the open-source GPUMD package, which is a general-purpose MD simulation software fully developed in CUDA language. In comparison to state-of-the-art MLPs, the NEP approach not only attains accuracy above the average but also demonstrates significantly higher computational efficiency [39]. It has been proven effective in modeling interatomic interactions across a broad range of materials under different temperatures and pressures, such as lithium-aluminum compounds [53], silicon [51,54-56] and carbon [57-60] systems, water systems [61], superatomic crystal [62], amorphous $HfO_2$ [63], alloy systems [44,46], and so on.

In the NEP framework [39,51,52], a feedforward neural network with a single hidden layer containing $N_{neu}$ neurons is used to model the site energy of atom $i$:

$$U_i = \sum_{\mu=1}^{N_{neu}} \omega_\mu^{(1)} \tanh\left(\sum_{\nu=1}^{N_{des}} \omega_{\mu\nu}^{(0)} q_\nu^i - b_\mu^{(0)}\right) - b^{(1)}, \tag{1}$$

where $\tanh(x)$ is the activation function in the hidden layer, $N_{des}$ is the descriptor components, $\omega^{(0)}$ is the connection weight matrix from the input layer (descriptor vector) to the hidden layer, $\omega^{(1)}$ is the connection weight vector from the hidden layer to the output node $U_i$, $b^{(0)}$ is the bias vector in the hidden layer, and $b^{(1)}$ is the bias for node $U_i$. The total number of parameters in the neural network is thus $(N_{des}+2)N_{neu}+1$.

For the MLPs, the descriptor is one of the most crucial aspects. In NEP, the descriptor vectors include radial descriptor components and angular descriptor components. The radial descriptor with $n_{max}^R + 1$ components is defined as:

$$q_n^i = \sum_{j \neq i} g_n(r_{ij}), \ 0 \leq n \leq n_{max}^R, \tag{2}$$

where the summation runs over all the neighbors of atom $i$ within a certain cutoff distance. The radial function $g_n(r_{ij})$ is defined as a linear combination of $N_{\text{bas}}^{\text{R}}+1$ basis functions $\{f_k(r_{ij})\}_{k=0}^{N_{\text{bas}}^{\text{R}}}$:

$$g_n(r_{ij}) = \sum_{k=0}^{N_{\text{bas}}^{\text{R}}} c_{nk}^{ij} f_k(r_{ij}), \qquad (3)$$

where $c_{nk}^{ij}$ are the expansion coefficients. The basis functions $f_k(r_{ij})$ are expressed as:

$$f_k(r_{ij}) = \frac{1}{2}[T_k(2(r_{ij}/r_c^{\text{R}}-1)^2 -1)+1]f_c(r_{ij}). \qquad (4)$$

In Eq. (2) and (3), both $n_{\max}^{\text{R}}$ and $N_{\text{bas}}^{\text{R}}$ are tunable hyperparameters. $T_k(x)$ is the $k^{\text{th}}$ order Chebyshev polynomial of the first kind. The cutoff function $f_c(r_{ij})$ is defined as follows:

$$f_c(r_{ij}) = \begin{cases} \frac{1}{2}[1+\cos(\pi\frac{r_{ij}}{r_c^{\text{R}}})], & r_{ij} \leq r_c^{\text{R}}; \\ 0, & r_{ij} > r_c^{\text{R}}. \end{cases} \qquad (5)$$

Here, $r_c^{\text{R}}$ is the cutoff distance of the radial descriptor components. Due to the summation over neighbors, the radial descriptor components defined above are invariant with respect to the permutation of atoms of the same type. The angular descriptor components include both radial and angular information. To keep a balance between accuracy and computational seed, we herein adopted the three-body $q_{nl}^i$ ( $0 \leq n \leq n_{\max}^{\text{A}}$, $0 \leq l \leq l_{\max}^{\text{3b}}$ ) and four-body $q_{nl_1l_2l_3}^i$ ( $0 \leq n \leq n_{\max}^{\text{A}}$, $1 \leq l_1 \leq l_2 \leq l_3 \leq l_{\max}^{\text{4b}}$ ) angular descriptor components to train the NEP model, and they can be defined as follows:

$$q_{nl}^i = \sum_{m=-l}^{l}(-1)^m A_{nlm}^i A_{nl(-m)}^i, \qquad (6)$$

$$q_{nl_1l_2l_3}^i = \sum_{m_1=-l_1}^{l_1}\sum_{m_2=-l_2}^{l_2}\sum_{m_3=-l_3}^{l_3}\begin{pmatrix} l_1 & l_2 & l_3 \\ m_1 & m_2 & m_3 \end{pmatrix} A_{nl_1m_1}^i A_{nl_2m_2}^i A_{nl_3m_3}^i, \qquad (7)$$

where the term $A_{nlm}^i$ is defined as:

$$A^i_{nlm} = \sum_{j \neq i} g_n(r_{ij}) Y_{lm}(\theta_{ij}, \phi_{ij}), \tag{8}$$

where $Y_{lm}(\theta_{ij}, \phi_{ij})$ are the spherical harmonics as a function of the polar angle $\theta_{ij}$ and the azimuthal angle $\phi_{ij}$ for the position vector $r_{ij} = \mathbf{r}_j - \mathbf{r}_i$ from atom $i$ to atom $j$.

In the training of NEP models, the separable natural evolution strategy (SNES) [64] is employed to optimize the free parameters **z** and minimize the loss function $L(\mathbf{z})$. The total loss function is defined as the sum of the weights of multiple loss functions:

$$L(\mathbf{z}) = \lambda_1 L_1 + \lambda_2 L_2 + \lambda_e \Delta U + \lambda_f \Delta F + \lambda_v \Delta W, \tag{9}$$

where $\Delta U$, $\Delta F$, $\Delta W$ are the root mean square errors (RMSEs) between reference and predicted values for energy, force, and virial, respectively. $L_1$ and $L_2$ represent the regularization terms and are proportional to the 1-norm and 2-norm of the training parameters, respectively. The weight factors $\lambda_1$, $\lambda_2$, $\lambda_e$, $\lambda_f$, and $\lambda_v$ for the independent loss function terms above are tunable hyperparameters.

## B. The NEP training hyperparameters

The NEP-TiAlNb model was trained using the GPUMD-v3.8 with the NEP4 model. The used training hyperparameters are listed in Supplemental Material Table S1 and described as follows. The cutoff radii for radial and angular descriptors are 6 Å and 4 Å, respectively, which are sufficient for most alloy systems. The Chebyshev polynomial expansion orders for both the radial and angular descriptor components are 8 Å. Both the radial and angular descriptors are constructed using 12 basis functions. The number of neurons in the hidden layer of the neural network is 60. The weighting factors in the loss function are $\lambda_1 = \lambda_2 = 0.05$, $\lambda_e = \lambda_f = 1.0$, and $\lambda_v = 0.5$. The population size was set to 80, and the training was executed with a batch size of 5000 structures for one million generations using the SNES algorithm. To achieve improved training accuracy, extra weight factors for force and shear virials were set to 1.0 and 10, respectively. The detailed descriptions of these hyperparameters are shown in Ref. [39,65].

## C. DFT calculations for training data

All DFT calculations were performed with the VASP code [66] using the Perdew-Burke-Ernzerhof (PBE) functional based on generalised gradient approximation and projector-augmented plane wave (PAW) methods [67] for the core region. The plane-wave cutoff energy was set to 600 eV and Γ-centered *k*-points with grid spacing of 0.15 Å$^{-1}$ were sampled in the Brillouin zone. The Methfessel-Paxton smearing method with order 1 and smearing width of 0.1 eV was used for partial electron occupancy. The energy threshold for self-consistency and the force threshold for structure relaxation were $10^{-6}$ eV and 0.02 eV/Å, respectively. Non-spin-polarized *ab initio* molecular dynamics (AIMD) simulations with supercell models were performed for extracting initial training structures. The ASE [68] and PyNEP [39] Python libraries were used for all structure manipulations and analysis of DFT results. The used INCAR file for static calculations is presented in Supplemental Material Note S1.

### D. Training data generation

The diverse training data covering a broad range of atomic local environments is critical for developing an effective and robust machine-learned interatomic potential. For the ternary NEP-TiAlNb model, the training data included the elemental, binary, and ternary systems. The detailed initial structure generation for each system is described as follows.

#### 1. Elemental systems (Ti, Al, Nb)

- Fully relaxed ground-state structures for elemental metals with HCP, BCC, and FCC crystals.
- Perturbation structures constructed by applying volume scaling from 0.94 to 1.06 with an interval of 0.01 and additional random perturbation with perturbation amplitudes of 0.01 Å and 0.03 times the cell length.
- Defect structures including surface, vacancy, and stacking faults configurations.
- NVT AIMD snapshots of the bulk and defect supercell structures. AIMD simulations were performed for 2000 steps with a timestep of 3 fs, and snapshots were extracted from each AIMD trajectory at intervals of 100 fs. The simulation temperatures increased gradually from 100 K to above the melting point in increments of 500 K.
- Elemental structures from the Materials Project and Open Quantum Materials Database.

#### 2. Binary systems (Ti-Al, Ti-Nb, Al-Nb)

- Solid solution structures constructed by partial substitution of 2×2×2 bulk supercells of one element with another, incorporating various crystals of HCP, BCC, and FCC. Twenty different concentrations of the form $A_xB_{1-x}$ were sampled with $x$ ranging from 0 to 100 at.%. For each concentration, atoms are randomly ordered and random perturbations were applied to both atom coordinates and cell vector, with the same perturbation amplitudes as those used in the elemental system.
- Defect structures including surface, vacancy, and stacking fault configurations for typical Ti-Al components ($\gamma$-TiAl, $\alpha_2$-Ti$_3$Al, D0$_{22}$-TiAl$_3$).
- NVT AIMD snapshots of bulk and defect structures for typical Ti-Al components at 100, 300, 1000, 2000, and 3000 K.
- Binary structures from the Materials Project and Open Quantum Materials Database.

#### 3. Ternary systems (Ti-Al-Nb)

- Solid solution structures constructed by 4×4×4 supercells with a random distribution of elements. Each solid solution structure was subjected to random perturbation with perturbation amplitudes of 0.01 Å and 0.03 times the cell length.
- Nb-doped structures constructed by random substitution of Ti or Al atoms in 3×3×3 bulk supercells of typical Ti-Al components. The maximum concentration of Nb in doped structures reaches 68.75 at.%.
- NVT AIMD snapshots of Nb-doped structures at 100, 300, 1000, 2000, and 3000 K.

- Ternary structures from the Materials Project and Open Quantum Materials Database.

Although the above initial training data has included diverse structures, the trajectories of extended MD simulations might go beyond the boundary of training data space, given that the NEP model is based on a neural net framework rather than a physical model. To further improve the transferability of the NEP model and the robustness of MD runs, we introduce an active learning strategy [39,62] by iteratively selecting candidate configurations from NEP-MD trajectories and training new NEP models based on the expanded dataset.

The NEP-MD trajectories used for sampling the candidate configurations cover a wide range of thermodynamic variables (temperature and pressure). Different bulk and defect configurations were used as MD input models, and then NEP-MD simulations with NVT or NPT ensemble were performed to explore the configuration space. The temperature of the MD simulations was systematically incremented within the range of 50-5000 K. In this manner, many structures far from equilibrium can be generated within the MD trajectories. Consequently, employing the farthest-point sampling method enables the selection of candidate structures that significantly differ from the initial structures, thereby forming a new training dataset.

## III. RESULTS AND DISCUSSIONS

### A. Workflow for the NEP-TiAlNb model

Figure 1(a) illustrates the workflow for fitting the NEP-TiAlNb model. Briefly, the workflow consists of three steps: constructing the initial training set, NEP train loops, and active learning to sample structures. The structures used to build the initial training set comprise three subclasses: elemental metals (Ti, Al, Nb), binary components of Ti-Al, Ti-Nb, and Nb-Al, as well as the ternary components Ti-Al-Nb. Initially, primitive cells of elemental metals with different lattices (FCC, BCC, HCP) are fully relaxed at 0 K using the VASP code (see Methods section for DFT calculation details). The relaxed bulk structures are then repeated to a 2×2×2 supercell, and a volume scaling from 0.94 to 1.06 with an interval of 0.01 is applied to induce uniform stretching or compression deformations. Subsequently, additional random perturbations are applied to the ion positions and supercell vectors of all initial crystal structures, with perturbation amplitudes of 0.01 Å and 0.03 times the cell length. Similar perturbations are employed for different concentration structures in binary components and ternary components. The initial training sets also include surface configurations and vacancy configurations. Additionally, stacking fault configurations are specifically constructed to model plastic deformation and the real dislocation behavior of TiAl alloys. To enhance the breadth of structures, the training dataset incorporates configurations from the Materials Project [69] and Open Quantum Materials Database [70]. For each bulk and defect supercell, snapshots are extracted from the AIMD trajectories at 100 fs intervals and further subjected to single-point calculations. Finally, atomic coordinates, total energy, forces, and virial tensors for each configuration are recorded to build the initial training set. More details on constructing the training sets are described in the Methods section.

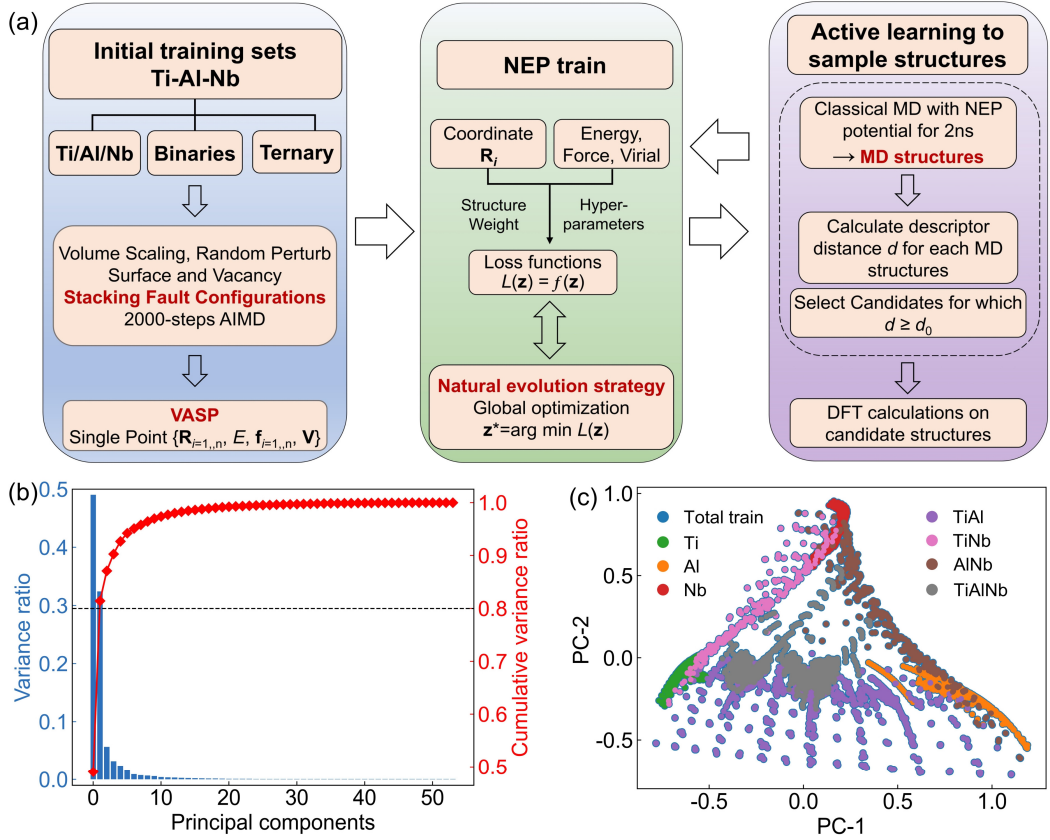

**FIG. 1.** (a) The workflow for fitting the NEP-TiAlNb model with an active learning scheme. $R_i$ denotes the atomic coordinate of atom $i$, $E$ is the total energy of each configuration, $V$ is the virial tensor of each configuration, $f_i$ is the force on atom $i$, and $n$ is the number of atoms in each configuration. (b) The explained variance ratio of the 54 principal components and cumulative variance ratio based on the NEP model. (c) Distribution of the full training data with 8524 structures in the two-dimensional PC subspace (spanned by PC-1 and PC-2) as reduced from the descriptor space.

In the NEP train loop step, initial training sets are input into the GPUMD package to train the initial NEP model. The SNES algorithm is adopted to optimize the free parameters and minimize the loss functions in NEP. After obtaining the NEP model, NVT and NPT simulations of bulk and defect structures are performed at selected temperatures of 50~5000 K using the LAMMPS [71] package. The active learning scheme based on the descriptor space of a pre-trained NEP model is used to select unsampled configurations from the above MD trajectories. These newly selected configurations have relatively large descriptor distances to existing configurations in the descriptor space and are then computed using DFT to form additional training datasets. Another train iteration is performed using all expanded datasets to build a new NEP model. After several train iterations, the active learning sampled structures can cover the setting MD processes and help to construct accurate and transferable MLPs (see Methods section for more details on active learning scheme). Note that the distance in the descriptor space provides a good quantitative uncertainty metric to be used in an active learning scheme as described in Ref. [39].

With the trained NEP model, we can calculate the descriptor space vectors for all structures in the training sets. To further visualize the distribution of training sets, principal component analysis (PCA) [72] is adopted to reduce the dimensionality of high-dimension descriptor space and capture the most significant variance in the data. The explained variance ratio of the 54 principal components and the cumulative variance ratio are shown in Fig. 1(b). The results indicate that the first two leading principal components combined explain at least 80% of the variance in the data, allowing us to visualize the training sets in the two-dimensional PC subspace. As depicted in Fig. 1(c), the overall distribution of the training set forms a triangular pattern in the two-dimensional space. The elemental structures are located at the three vertices of the triangle, binary systems are respectively distributed along the three sides formed by connecting the constituent elements, and ternary compounds are positioned within the interior of the triangle. The distribution features mentioned above satisfy the concept of chemical generalizability [73], implying that various $n$-component ($n \geq 3$) structures fall comfortably within the space spanned by the 1-component and 2-component training structures.

## B. Potential training and validation

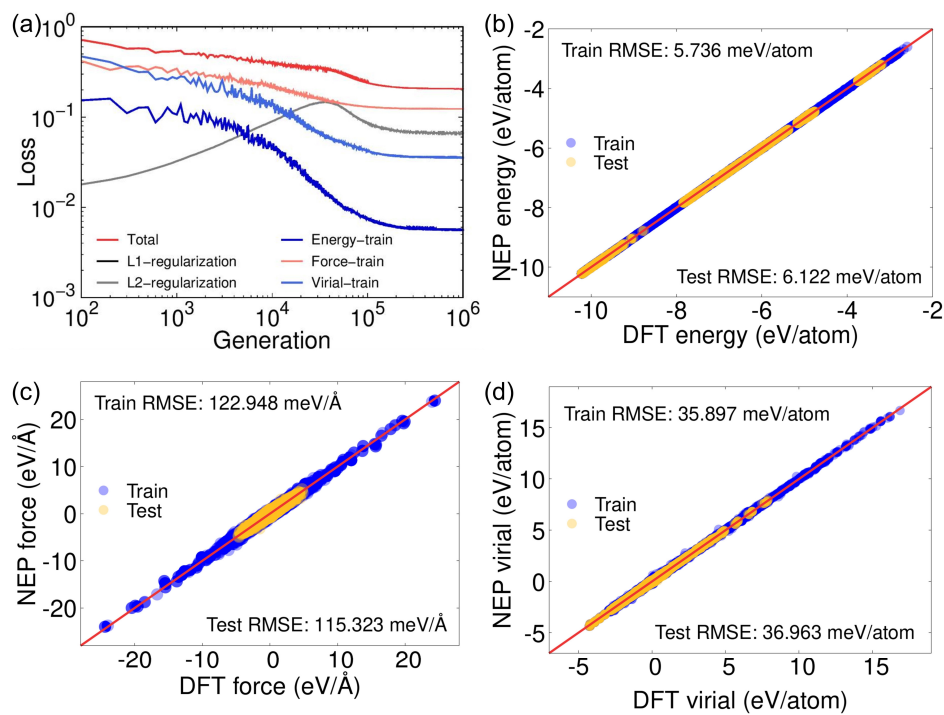

**FIG. 2.** Performance of the NEP-TiAlNb model. (a) The evolution of various loss functions as a function of generations. Parity plots of NEP calculated (b) energy, (c) force, and (d) virial as compared to the DFT reference values for the training and testing datasets.

Table S2 in Supplemental Material summarizes the number of datasets for each system. The number of structures in the total training set is 8524. A test set is further generated to validate the generalizability of the fitted NEP model, and the test set to training set ratio is 1:9. Figure 2(a) shows the evolution of loss functions for various components as a function of train generations. The training has been performed for $1\times10^6$ generations until all loss functions converged, after which the calculated energy [Fig. 2(b)], force [Fig. 2(c)] and virial [Fig. 2(d)] of the NEP model are compared with the DFT reference values. The overall training and test RMSE for energy, force, and virial are within 6.122 meV/atom, 122.948 meV/Å, and 36.963 meV/atom, respectively. This indicates that the trained NEP-TiAlNb model exhibits high accuracy comparable to the DFT calculations. Meanwhile, the energy, force, and virial in the dataset are in a wide range, ensuring the stability of long-time MD simulations based on the trained NEP model in various conditions.

## C. Basic physical properties

### Equations of state

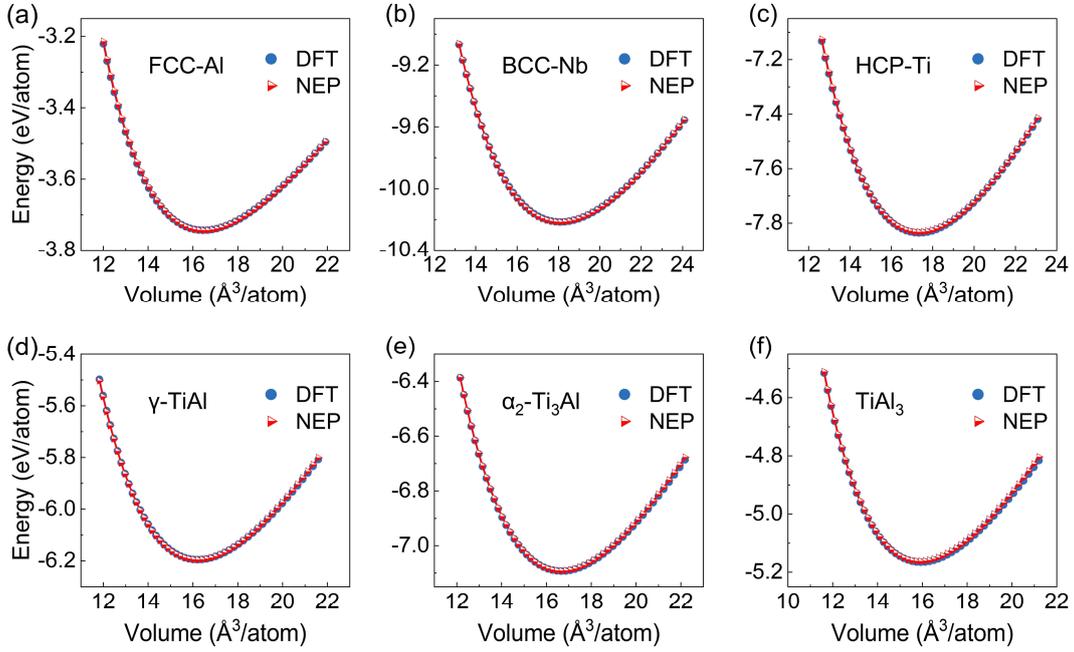

**FIG. 3.** The equations of state for (a) FCC-Al, (b) BCC-Nb, (c) HCP-Ti, (d) γ-TiAl, (e) $\alpha_2$-Ti$_3$Al and (f) D0$_{22}$-TiAl$_3$ calculated by NEP and DFT.

We then shift to calculate the basic physical properties of typical phases to examine the accuracy of the developed NEP-TiAlNb model. Figure 3 shows the equations of state (EOS) curves for six typical structures (FCC-Al [Fig. 3(a)], BCC-Nb [Fig. 3(b)], HCP-Ti [Fig. 3(c)], γ-TiAl [Fig. 3(d)], $\alpha_2$-Ti$_3$Al [Fig. 3(e)], and D0$_{22}$-TiAl$_3$ [Fig. 3(f)]) calculated by NEP and DFT. For each point on the EOS curve, a 2×2×2 supercell is scaled to the desired volume and equilibrated in DFT (i.e., the supercell volume is fixed but the supercell shape and ion positions are allowed to relax). For consistency, the energy and volume per atom are predicted by NEP and DFT with the same equilibrated supercell. For all six structures, the NEP and DFT EOS are in excellent agreement over the entire volume range examined (corresponding to ±10% volumetric strain). The RMSE between NEP and DFT is smaller than 1.5 meV/atom. The accurate EOS curves given by NEP are important for predicting the mechanical response of materials under a wide range of volumetric strains and pressures.

**TABLE I.** Lattice parameters $a$, $c$, $c/a$, atomization energy $E_{am}$, equilibrium volume per atom $V_0$, and energy differences $\Delta E$ for elemental structures from experiment (Expt.), DFT, NEP, EAM and MEAM. The percentage errors with respect to experimental or DFT (when experimental data are unavailable) values are indicated in the parenthesis.

| Structure | Property | Expt. | DFT | NEP | EAM(MEAM) |
|---|---|---|---|---|---|
| FCC-Al | $a$ (Å) | 4.046[a] | 4.039 | 4.040 (0.15%) | 4.052 (0.15%) |
| | $E_{am}$ (eV/atom) | -- | -3.745 | -3.745 (0%) | -3.353 |
| | $V_0$ (Å$^3$/atom) | -- | 16.53 | 16.53 (0%) | 16.61 (0.5%) |
| | $\Delta E_{bcc-fcc}$ (meV/atom) | -- | 89 | 91 (2.2%) | 148 (64.8%) |
| | $\Delta E_{hcp-fcc}$ (meV/atom) | -- | 11 | 10 (9%) | 16 (50%) |
| BCC-Nb | $a$ (Å) | 3.303[b] | 3.307 | 3.307 (0.12%) | 3.308 (0.15%) |
| | $E_{am}$ (eV/atom) | -- | -10.217 | -10.217 (0%) | -7.091 |
| | $V_0$ (Å$^3$/atom) | -- | 18.092 | 18.092 (0%) | 18.097 (0.03%) |
| | $\Delta E_{fcc-bcc}$ (meV/atom) | -- | 322 | 325 (0.9%) | 187 (41.5%) |
| | $\Delta E_{hcp-bcc}$ (meV/atom) | -- | 286 | 289 (1%) | 187 (34.3%) |
| HCP-Ti | $a$ (Å) | 2.947[c] | 2.936 | 2.934 (0.4%) | 2.930 (0.6%) |
| | $c$ (Å) | 4.674[c] | 4.648 | 4.654 (0.4%) | 4.676 (0.04%) |
| | $c/a$ | 1.586[c] | 1.583 | 1.586 (0%) | 1.596 (0.6%) |
| | $E_{am}$ (eV/atom) | -- | -7.834 | -7.833 (0.01%) | -4.831 |
| | $V_0$ (Å$^3$/atom) | -- | 17.337 | 17.335 (0.01%) | 17.396 (0.3%) |
| | $\Delta E_{fcc-hcp}$ (meV/atom) | -- | 58 | 55 (5.2%) | 39 (34.5%) |
| | $\Delta E_{bcc-hcp}$ (meV/atom) | -- | 110 | 108 (1.8%) | 111 (0.9%) |

[a]Ref. [74]; [b]Ref. [75]; [c]Ref. [76,77].

Based on the calculated EOS curves, the physical properties, including lattice constants ($a$, $c$, $c/a$), atomization energy $E_{am}$, equilibrium volume $V_0$, and energy differences $\Delta E$, are summarized in Table I for pure metals and Table II for binary systems, respectively. For pure metals, we compare the accuracy of the NEP model with state-of-the-art classical MEAM potentials for Al [78] and Ti [79], and EAM potential for Nb [80]. In the case of binary TiAl system, three widely used classical potentials, namely EAM-Zope [22], EAM-Farkas [23], and MEAM-Kim [24], are used for comparison. Additionally, the corresponding DFT and available experimental data are also listed as reference values. We quantify the deviations of the NEP model using the absolute value of relative error with respect to the experiment or DFT data (when experimental values are unavailable). The NEP model accurately reproduces the lattice constants, equilibrium volume and energies for both elemental and binary systems in excellent agreement with the reference values. The differences between DFT and the NEP model are smaller than 0.006 Å, 0.002 Å$^3$/atom, and 3 meV/atom for lattice constants, equilibrium volume, and energies, respectively. The EAM and MEAM potentials also have accurate lattice parameters for three pure metals within a 1% deviation from the experimental results. In terms of $\Delta E$ for pure metals, the maximum deviation of NEP model is within 10%, which is more accurate than that of ~65% in MEAM potential.

**TABLE II.** Lattice parameters $a$, $c$, $c/a$, atomization energy $E_{am}$, equilibrium volume per atom $V_0$, energy differences $\Delta E$ for typical binary TiAl structures from experiment, DFT, NEP, EAM and MEAM. The percentage errors with respect to experimental or DFT (when experimental data are unavailable) values are indicated in the parenthesis.

| Structure | Property | Expt. | DFT | NEP | EAM-Zope | EAM-Farkas | MEAM-KIM |
|---|---|---|---|---|---|---|---|
| γ-TiAl | $a$ (Å) | 3.988[a] | 3.987 | 3.986 (0.05%) | 3.998 (0.25%) | 3.906 (2.05%) | 4.018 (0.75%) |
|  | $c$ (Å) | 4.067[a] | 4.078 | 4.084 (0.4%) | 4.187 (3%) | 4.152 (2.1%) | 4.099 (0.8%) |
|  | $c/a$ | 1.021[a] | 1.023 | 1.024 (0.3%) | 1.047 (2.5%) | 1.063 (4.1%) | 1.020 (0.1%) |
|  | $E_{am}$ (eV/atom) | -- | -6.194 | -6.195 (0.02%) | -4.508 | -4.399 | -4.501 |
|  | $V_0$ (Å³/atom) | -- | 16.222 | 16.222 (0%) | 16.728 (3.1%) | 15.840 (2.4%) | 16.548 (2%) |
| α₂-Ti₃Al | $a$ (Å) | 5.81[b] | 5.754 | 5.752 (0.9%) | 5.784 (0.4%) | 5.730 (1.4%) | 5.805 (0.08%) |
|  | $c$ (Å) | 4.65[b] | 4.656 | 4.654 (0.08%) | 4.750 (2.2%) | 4.662 (0.3%) | 4.655 (0.1%) |
|  | $c/a$ | 0.801[b] | 0.809 | 0.809 (1%) | 0.821 (2.5%) | 0.813 (1.5%) | 0.802 (0.1%) |
|  | $E_{am}$ (eV/atom) | -- | -7.092 | -7.091 (0.01%) | -4.766 | -4.676 | -4.776 |
|  | $V_0$ (Å³/atom) | -- | 16.664 | 16.666 (0.01%) | 17.201 (3.2%) | 16.568 (0.6%) | 16.976 (1.9%) |
| D0₂₂-TiAl₃ | $a$ (Å) | 3.854[c] | 3.843 | 3.841 (0.3%) | 4.049 (5.1%) | 3.939 (2.2%) | 4.037 (4.7%) |
|  | $c$ (Å) | 8.584[c] | 8.621 | 8.620 (0.4%) | 8.139 (5.2%) | 8.055 (6.2%) | 8.049 (6.2%) |
|  | $c/a$ | 2.227[c] | 2.243 | 2.244 (0.8%) | 2.010 (9.7%) | 2.045 (8.2%) | 1.994 (10.5%) |
|  | $E_{am}$ (eV/atom) | -- | -5.164 | -5.163 (0.02%) | -4.021 | -3.877 | -4.065 |
|  | $V_0$ (Å³/atom) | -- | 15.931 | 15.931 (0%) | 16.680 (4.7%) | 15.620 (2%) | 16.402 (3%) |

[a]Ref. [81]; [b]Ref. [82]; [b]Ref. [83].

For the complex TiAl intermetallic compounds (Table II), only the NEP model has the best overall agreement with experimental values with deviations within 1%. The EAM-Zope potential yields the accurate lattice parameter $a$ in both the γ and α₂ phases (within 0.5%), while its $c/a$ ratios are ~2.5% larger than experimental values. The EAM-Farkas potential underestimates $a$ in all three Ti-Al binary phases (1.4-2.2%). Among these three classical potentials, the MEAM-Kim potential has the best accurate lattice parameters agreement with experimental values for both the γ and α₂ phases with errors within 1%. Compared to the NEP model, both the EAM and MEAM potentials exhibit significantly poorer accuracy in describing the equilibrium parameters of the D0₂₂-TiAl₃ phase, with errors of up to 11%. Note that the $E_{am}$ calculated by EAM and MEAM potentials differ significantly from DFT values, which is attributed to the fact that classical potentials use the experimental cohesion energies parameters for the fitting.

**Phonon dispersion**

As shown in Fig. 4, the NEP calculated phonon dispersion curves of six phases are compared to EAM(MEAM), DFT, and available experimental data to capture the lattice vibration behavior. The PHONOPY [84] package with a finite-displacement method is used to compute the phonon dispersion along high-symmetry directions in the Brillouin zone. For FCC-Al [Fig. 4(a)], the NEP results match very well with DFT and experiment data [85]. However, the MEAM potential fails to reproduce the vibrational properties at short wavelengths, particularly the zone boundary phonons, in contrast to NEP. For BCC-Nb [Fig. 4(b)], our NEP results match the experiment [86] more closely than the EAM results over much of the Brillouin zone. The EAM results also exhibit well in the low-frequency region but have poor agreement at the zone boundaries H and N. For the HCP-Ti phase [Fig. 4(c)], both NEP and MEAM exhibit good agreement with the overall trend in the experimental data [79], especially for acoustic phonons. Besides, the NEP and MEAM also overestimate the frequencies of optical phonons at Γ and underestimate the optical phonons at K and M. Notably, the deep potential developed for Ti also exhibits similar deviations in phonon dispersions within the HCP phase [43]. In terms of more complex binary TiAl systems [Fig. 4(d-f)], the phonon dispersions predicted by NEP are generally consistent with those of the DFT calculation. The deviation of the EAM phonon dispersions lies in the high-frequency region when compared to the DFT results. Overall, the NEP exhibits excellent performance in describing the phonon dispersions of both elemental and binary systems.

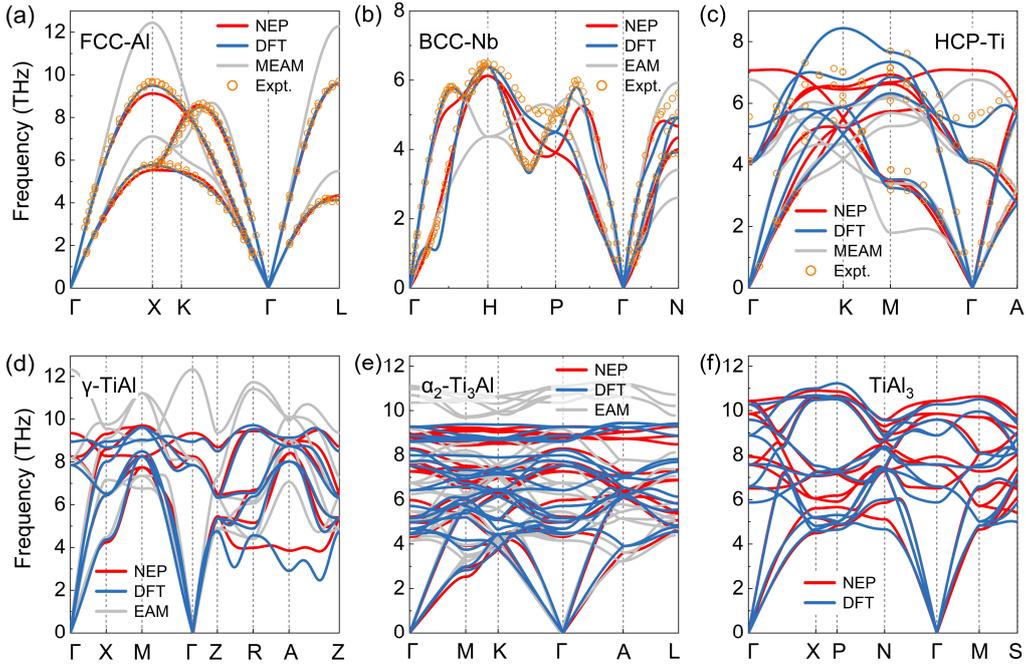

**FIG. 4.** Phonon dispersion curves of (a) FCC-Al, (b) BCC-Nb, (c) HCP-Ti, (d) γ-TiAl, (e) α$_2$-Ti$_3$Al and (f) D0$_{22}$-TiAl$_3$ at 0 K. The experimental data are taken from [79,85,86]. Note that the EAM phonon dispersions of D0$_{22}$-TiAl$_3$ are not present here due to the unreasonable results.

**0 K elastic constants**

Tables III and IV present the 0 K elastic constants calculated by NEP and classical potentials in comparison with DFT and available experiment data. As summarized in Table III, the NEP model accurately predicts elastic constants close to the target DFT values for elemental systems. For FCC-Al, the overall deviation predicted by the NEP model is comparable to that of MEAM potential (within 10%). For BCC-Nb, while the NEP model underestimates $C_{44}$ by approximately 48% compared to experimental values, it is closer to the DFT value. For elemental Ti, the NEP model accurately reproduces the elastic constants of three crystals, including the HCP, BCC, and FCC phases (see Table S3 in Supplemental Material for results of BCC and FCC). The largest deviation for HCP-Ti occurs in $C_{66}$, with an underestimation of ~29%. Overall, the classical EAM or MEAM potentials offer better accuracies in 0 K elastic constants for pure metals, particularly in comparison to experimental values, which is not surprising since the basic experimental properties have been used to tune the parameters of these force fields. Meanwhile, the classical potentials of Nb and Ti used here for comparison encompass only a single element; if multiple elements are included, the description of elemental properties could potentially deteriorate, as observed in EAM-Farkas [50].

**TABLE III.** Elastic constants $C_{ij}$, bulk modulus $B_v$ (voigt), shear modulus $G_v$ (voigt), for elemental structures from experiment, DFT, NEP, EAM and MEAM. The percentage errors with respect to experimental values are indicated in the parenthesis.

| Structure | Property | Expt. | DFT | NEP | EAM(MEAM) |
|---|---|---|---|---|---|
| FCC-Al | $C_{11}$ (GPa) | 114.3[a] | 113.2 | 118.5 (3.6%) | 110.5 (3.3%) |
| | $C_{12}$ (GPa) | 61.9[a] | 62.3 | 57.6 (6.9%) | 61 (1.5%) |
| | $C_{44}$ (GPa) | 31.6[a] | 32.9 | 32.5 (2.8%) | 28.4 (10%) |
| | $B_v$ (GPa) | 79.4[a] | 79.2 | 78.2 (1.5%) | 77.5 (2.4%) |
| | $G_v$ (GPa) | 29.4[a] | 29.9 | 31.8 (8.2%) | 27 (8.2%) |
| BCC-Nb | $C_{11}$ (GPa) | 246.5[b] | 247.8 | 230.8 (6.4%) | 233.1 (5.4%) |
| | $C_{12}$ (GPa) | 134.5[b] | 133 | 139.3 (3.6%) | 123.8 (8%) |
| | $C_{44}$ (GPa) | 28.7[b] | 17.2 | 14.9 (48%) | 32.1 (11.8%) |
| | $B_v$ (GPa) | 173[b] | 172 | 170.2 (1.6%) | 160.3 (7.3%) |
| | $G_v$ (GPa) | 38[b] | 30 | 27.1 (28.7%) | 41.1 (8.2%) |
| HCP-Ti | $C_{11}$ (GPa) | 176.1[c] | 171.4 | 156.6 (11%) | 174.2 (1%) |
| | $C_{12}$ (GPa) | 86.9[c] | 93.2 | 93.8 (8%) | 94.7 (9%) |
| | $C_{13}$ (GPa) | 68.3[c] | 74.6 | 63.5 (7%) | 72.3 (5.8%) |
| | $C_{33}$ (GPa) | 190.5[c] | 197.2 | 215.5 (13%) | 187.9 (1.4%) |
| | $C_{44}$ (GPa) | 50.8[c] | 42.7 | 43.2 (15%) | 57.7 (13.6%) |
| | $C_{66}$ (GPa) | 44.3[c] | 44.4 | 31.4 (29%) | 39.7 (10.4%) |
| | $B_v$ (GPa) | 109.9[c] | 113.8 | 107.8 (2%) | 112.8 (2.6%) |
| | $G_v$ (GPa) | 50.5[c] | 45.8 | 44.1 (12.6%) | 50.8 (0.6%) |

[a]Ref. [87]; [b]Ref. [88]; [c]Ref. [76].

Table IV shows the 0 K elastic constants of three typical Ti-Al binary structures and all of them have six independent elastic constants. The NEP calculated elastic constants match well with corresponding DFT and experimental results for all binary structures. For γ-TiAl and $α_2$-Ti$_3$Al, the deviations of NEP are less than ~6% from DFT and ~15% from experiment data (similar deviations exist between DFT and experiment). For $D0_{22}$-TiAl$_3$ phase, substantial disparities between DFT and experimental values lead to a 76% overestimation of the NEP $C_{12}$ constant, while the remaining elastic constants are within 17% deviations from experimental values. More importantly, only the NEP model reproduces the negative Cauchy pressure $C_{13}$-$C_{44}$ = -28.3 GPa in γ-TiAl, which is close to the DFT value of -25 GPa. The negative Cauchy pressure arises from the valence *sp* electrons and has been challenging to be reproduced by classical interatomic potentials [89]. On the other hand, all the classical potentials have substantial differences exceeding 50% in 0 K elastic constants when compared to the experimental values. Overall, the NEP model shows significant improvement in elastic constants for both elemental systems and complex TiAl intermetallic compounds, making it more suitable for predicting the mechanical properties of TiAl systems.

**Table IV.** Elastic constants $C_{ij}$, bulk modulus $B_v$ (voigt), shear modulus $G_v$ (voigt), for typical Ti-Al binary structures from experiment, DFT, NEP, EAM and MEAM. The percentage errors with respect to experimental values are indicated in the parenthesis.

| Structure | Property | Expt. | DFT | NEP | EAM-Zope | EAM-Farkas | MEAM-Kim |
|---|---|---|---|---|---|---|---|
| γ-TiAl | $C_{11}$ (GPa) | 187[a] | 170 | 175.2 (6.3%) | 195.8 (4.7%) | 246 (31.6%) | 181.2 (3.1%) |
| | $C_{12}$ (GPa) | 74.8[a] | 87.7 | 86.1 (15%) | 107 (43%) | 118.2 (58%) | 76.4 (2.1%) |
| | $C_{13}$ (GPa) | 74.8[a] | 86.1 | 85.2 (14%) | 113.9 (52.3%) | 190.6 (154.8%) | 134 (79%) |
| | $C_{33}$ (GPa) | 182[a] | 164.2 | 162.4 (10.8%) | 213.3 (17.2%) | 352 (93.4%) | 234.4 (28.8%) |
| | $C_{44}$ (GPa) | 109[a] | 111.1 | 113.5 (4.1%) | 92.1 (15.5%) | 146.1 (34%) | 86 (21%) |
| | $C_{66}$ (GPa) | 81.2[a] | 73.7 | 73.5 (9.5%) | 84.7 (4.3%) | 71.5 (12%) | 62 (23.6%) |
| | $C_{12}$-$C_{66}$ (GPa) | -6.4[a] | 14.0 | 12.6 | 22.3 | 45 | 14.4 |
| | $C_{13}$-$C_{44}$ (GPa) | -34.2[a] | -25.0 | -28.3 | 22 | 44 | 48 |
| | $B_v$ (GPa) | 111.6[a] | 113.7 | 114.0 (2.2%) | 141.6 (26.8%) | 204.8 (83.5%) | 142.8 (28%) |
| | $G_v$ (GPa) | 82[a] | 75.5 | 77.2 (5.8%) | 71.8 (12.4%) | 95.7 (16.7%) | 63.6 (22.4%) |
| α2-Ti3Al | $C_{11}$ (GPa) | 183[b] | 182 | 181 (1.1%) | 199 (8.7%) | 233.6 (27.6%) | 200.7 (9.7%) |
| | $C_{12}$ (GPa) | 89.1[b] | 90 | 94.9 (6.5%) | 89 (0.1%) | 109.3 (22.6%) | 107.8 (21%) |
| | $C_{13}$ (GPa) | 62.6[b] | 60.1 | 60.2 (3.8%) | 74.3 (18.7%) | 88.7 (41.7%) | 91.3 (45.8%) |
| | $C_{33}$ (GPa) | 225[b] | 238 | 240.8 (7%) | 224.9 (0.04%) | 285.5 (26.8%) | 239 (6.2%) |
| | $C_{44}$ (GPa) | 64.1[b] | 54.0 | 56.5 (12%) | 51.2 (20.1%) | 57.8 (9.8%) | 45.5 (29%) |
| | $C_{66}$ (GPa) | 47[b] | 42 | 44.8 (4.6%) | 55 (17%) | 62.2 (32.3%) | 46.4 (1.3%) |
| | $C_{12}$-$C_{66}$ (GPa) | 42.1[b] | 48 | 50.1 | 34 | 47.1 | 62.1 |
| | $C_{13}$-$C_{44}$ (GPa) | -1.5[b] | 6.1 | 3.7 | 23.1 | 31 | 46 |
| | $B_v$ (GPa) | 113.3[b] | 113.6 | 114.8 (0.2%) | 122 (7.7%) | 147.4 (30%) | 135.7 (19.8%) |
| | $G_v$ (GPa) | 60.2[b] | 56.1 | 57.4 (4.6%) | 57.2 (5%) | 66.7 (10.8%) | 50.8 (15.6%) |
| D0$_{22}$-TiAl3 | $C_{11}$ (GPa) | 217.7[c] | 196 | 187.6 (13.8%) | 169.4 (22.2%) | 264.2 (21.4%) | 169.8 (22%) |
| | $C_{12}$ (GPa) | 57.7[c] | 87 | 101.8 (76.4%) | 98.8 (71.2%) | 187.9 (225.6%) | 115.1 (99.5%) |
| | $C_{13}$ (GPa) | 45.5[c] | 47 | 38.9 (14.5%) | 89.2 (96%) | 157.3 (245.7%) | 95.7 (110.3%) |
| | $C_{33}$ (GPa) | 217.5[c] | 220 | 222.3 (2.2%) | 139.6 (35.8%) | 232.6 (7%) | 190.4 (12.5%) |
| | $C_{44}$ (GPa) | 92[c] | 95 | 76.5 (16.8%) | 62.2 (32.4%) | 81.6 (11.3%) | 60.4 (34.3) |
| | $C_{66}$ (GPa) | 116.5[c] | 129 | 129.6 (11.2%) | 71 (39%) | 116.6 (0.1%) | 63 (46%) |
| | $B_v$ (GPa) | 105.6[c] | 108.2 | 106.3 (0.6%) | 114.7 (8.6%) | 196.2 (85.8%) | 126.9 (20.2%) |
| | $G_v$ (GPa) | 93.7[c] | 92.5 | 84.4 (10%) | 52.5 (44%) | 73.2 (21.8%) | 51.7 (44.8%) |

[a]Ref. [90]; [b]Ref. [82]; [c]Ref. [91].

**Surface energies and vacancy formation energies**

Our attention now turns to evaluating the defect properties of TiAl systems. Tables V and VI provide insights into the low-index surface energies and vacancy formation energies for pure metals and binary TiAl structures, respectively. For elemental structures, the NEP model performs exceptionally well in surface energies within 4% errors from DFT values. The largest discrepancy lies in the basal plane of HCP-Ti, with an underestimation of 0.07 J/m$^2$. In contrast, the MEAM potential overestimates the {100} and {110} surface energies of FCC-Al by ~22% and 18%, respectively. For HCP-Ti, the MEAM potential significantly underestimates surface energies for all four planes with deviations exceeding 18%. The EAM potential for Nb exhibits well results in surface energies within 5% errors from DFT values. The vacancy formation energy $E_{vf}$ given by the NEP model also matches well with the DFT values, especially for FCC-Al (within 2%). The $E_{vf}$ of BCC-Nb in NEP is ~11.5% lower than the DFT value, consistent with the deviation observed with the EAM potential. For $E_{vf}$ in HCP-Ti, the accuracy of MEAM is slightly higher than that of the NEP model.

**Table V.** Relaxed surface energies $\sigma$, vacancy formation energies $E_{vf}$ for elemental structures from DFT, NEP, EAM and MEAM. The percentage errors with respect to DFT values are indicated in the parenthesis.

| Structure | Property | DFT | NEP | EAM (MEAM) |
|---|---|---|---|---|
| FCC-Al | $\sigma_{\{100\}}$ (J/m$^2$) | 0.890 | 0.920 (3.4%) | 1.088 (22.2%) |
| | $\sigma_{\{110\}}$ (J/m$^2$) | 0.960 | 0.974 (1.4%) | 1.135 (18.2%) |
| | $\sigma_{\{111\}}$ (J/m$^2$) | 0.810 | 0.801 (1.1%) | 0.752 (7.2%) |
| | $E_{vf}$ (eV) | 0.64 | 0.65 (1.6%) | 0.67 (4.7%) |
| BCC-Nb | $\sigma_{\{100\}}$ (J/m$^2$) | 2.31 | 2.32 (0.4%) | 2.36 (2.2%) |
| | $\sigma_{\{110\}}$ (J/m$^2$) | 2.05 | 2.01 (2%) | 2.04 (0.5%) |
| | $\sigma_{\{111\}}$ (J/m$^2$) | 2.35 | 2.38 (1.3%) | 2.47 (5.1%) |
| | $E_{vf}$ (eV) | 2.78 | 2.46 (11.5%) | 3.10 (11.5%) |
| HCP-Ti | $\sigma_{basal}$ (J/m$^2$) | 1.95 | 1.87 (4%) | 1.47 (24.6%) |
| | $\sigma_{prism}$ (J/m$^2$) | 2.02 | 1.96 (2.9%) | 1.55 (23.3%) |
| | $\sigma_{pyr.I}$ (J/m$^2$) | 1.91 | 1.94 (1.6%) | 1.52 (20.4%) |
| | $\sigma_{pyr.II}$ (eV) | 2.05 | 2.03 (0.9%) | 1.68 (18%) |
| | $E_{vf}$ (eV) | 2.06 | 2.22 (7.8%) | 2.19 (6.3%) |

In typical binary Ti-Al structures (Table VI), the NEP calculated surface energies $\sigma$ are within 3% errors from the DFT values, demonstrating greater accuracy compared to classical potentials. For γ and α$_2$ phases, the EAM potentials (EAM-Zope and EAM-Farkas) underestimate $\sigma$ more than 30% from DFT values. Among the classical potentials, the MEAM-Kim potential provides the most accurate prediction for $\sigma$, with a deviation of within 14% from the DFT values. The NEP also accurately reproduces the vacancy formation energies for both Ti vacancies and Al vacancies in the TiAl binary structures. For γ-TiAl, the NEP has its $E_{vf}$ within ~1% (Ti vacancies) and ~6% (Al vacancies) from DFT values. However, all classical potentials show substantial deviations exceeding 25% for $E_{vf-Al}$. In the case of α$_2$-Ti$_3$Al, the NEP calculated $E_{vf-Ti}$ and $E_{vf-Al}$ closely match the DFT values with deviations ~1% and 2%, respectively. The three classical potentials substantially underestimate the vacancy formation energy by more than 15% ($E_{vf-Ti}$) and 25% ($E_{vf-Al}$), respectively. The largest discrepancy within the NEP model is ~24% in $E_{vf-Ti}$ of D0$_{22}$-TiAl$_3$, whereas EAM-Zope and MEAM-Kim outperform NEP in this aspect. For $E_{vf-Al}$ in TiAl$_3$, the NEP slightly underestimates by around 13%, while classical potentials exhibit larger errors exceeding 28%. Overall, the NEP model shows more accurate results than classical potentials in describing the defect properties of elemental and binary systems.

**Table VI.** Relaxed surface energies $\sigma$, vacancy formation energies $E_{vf-Ti}$ (Ti vacancies) and $E_{vf-Al}$ (Al vacancies) for binary Ti-Al structures from DFT, NEP, EAM and MEAM. The percentage errors with respect to DFT values are indicated in the parenthesis.

| Structure | Property | DFT | NEP | EAM-Zope | EAM-Farkas | MEAM-Kim |
|---|---|---|---|---|---|---|
| γ-TiAl | $\sigma_{\{111\}}$ (J/m$^2$) | 1.71 | 1.68 (1.7%) | 1.15 (32.7%) | 1.17 (31.6%) | 1.48 (13.5%) |
| | $E_{vf-Ti}$ (J/m$^2$) | 1.762 | 1.779 (0.9%) | 2.022 (14.8%) | 1.658 (6%) | 1.733 (1.6%) |
| | $E_{vf-Al}$ (J/m$^2$) | 2.712 | 2.547 (6.1%) | 1.969 (27.4%) | 2.005 (26.1%) | 2.014 (25.7%) |
| α$_2$-Ti$_3$Al | $\sigma_{\{0001\}}$ (J/m$^2$) | 1.96 | 1.91 (2.6%) | 1.25 (36.2%) | 0.98 (50%) | 1.88 (4.1%) |
| | $E_{vf-Ti}$ (J/m$^2$) | 2.194 | 2.181 (0.6%) | 1.881 (14.3%) | 1.437 (34.5%) | 1.739 (20.7%) |
| | $E_{vf-Al}$ (J/m$^2$) | 3.652 | 3.582 (1.9%) | 2.712 (25.7%) | 2.654 (27.3%) | 2.471 (32.3%) |
| D0$_{22}$-TiAl$_3$ | $E_{vf-Ti}$ (J/m$^2$) | 2.382 | 2.962 (24.3%) | 2.048 (14%) | 1.701 (28.6%) | 1.967 (17.4%) |
| | $E_{vf-Al}$ (J/m$^2$) | 2.336 | 2.034 (13%) | 1.321 (43.5%) | 1.682 (28%) | 1.490 (36.2%) |

**γ-Line and γ-Surface**

TiAl intermetallic compounds exhibit complex slip systems involving super-dislocations and corrugated slip planes. Superdislocations possess large Burgers vectors, often dissociating into partial dislocations with smaller Burgers vectors. The processes of dislocation dissociation, nucleation, and glide behavior are strongly influenced by the generalized stacking fault energies, noted as γ-line and γ-surface on each slip plane. Therefore, accurate stacking fault energies are crucial for modeling of dislocation behavior and plasticity in intermetallic structures. We then further examine the stacking fault energies for γ-TiAl and α$_2$-Ti$_3$Al.

Dislocation glide in γ-TiAl occurs on the {111} close-packed planes, and three different stacking faults are relevant for dislocation core dissociations, including intrinsic stacking fault (SISF), antiphase boundary (APB), and complex stacking fault (CSF). To obtain the γ-lines and γ-surface of γ-TiAl, the sliding vector $\boldsymbol{b}$ identifying the tilt of the supercell is given by a linear combination of two orthogonal basic vectors of $[\bar{1}10]/2$ and $[11\bar{2}]/2$:

$$\boldsymbol{b} = u[\bar{1}10]/2 + v[11\bar{2}]/2, \quad (10)$$

where $u$ and $v$ represent the fractional coordinates of two basic vectors. The corresponding sliding vector for SISF, APB and CSF are expressed as:

$$\boldsymbol{b}_{\text{SISF}} = \frac{1}{3}\left(\frac{1}{2}[11\bar{2}]\right),$$

$$\boldsymbol{b}_{\text{APB}} = \frac{1}{2}\left(\frac{1}{2}[\bar{1}10]\right) + \frac{1}{2}\left(\frac{1}{2}[11\bar{2}]\right), \quad (11)$$

$$\boldsymbol{b}_{\text{CSF}} = \frac{1}{2}\left(\frac{1}{2}[\bar{1}10]\right) + \frac{5}{6}\left(\frac{1}{2}[11\bar{2}]\right).$$

The CSF governs the ordinary <$\bar{1}$10]/2 dislocation, while the SISF and APB control the <$\bar{1}$10] and <11$\bar{2}$]/2 super-dislocations. Figure 5 shows the stacking fault energies on the {111} plane of γ-TiAl, as determined by NEP, EAM-Zope, and DFT. The NEP accurately reproduces the DFT γ-lines for slipping along [11$\bar{2}$] [Fig. 5(a) and (b)] and [$\bar{1}$10] [Fig. 5(c)] directions. However, the EAM-Zope potential substantially underestimates the overall γ-lines from those calculated by DFT. Most critically, for γ-lines along [11$\bar{2}$] direction [Fig. 5(a)], the NEP gives accurate SISF energy within 10% error from DFT values, while the three classical potentials substantially underestimate SISF exceeding 50% (Table VII). In addition, the MEAM-Kim potential even gives a negative SISF energy of ~-46 mJ/m$^2$, with a deviation of ~125%. The NEP reproduces the APB and CSF energies differing by ~0.3% and 6.2 % from the DFT values, respectively. Yet, the classical potentials underestimate APB and CSF energies by over 11% (Table VII). The DFT and NEP calculated APB energies are unstable with a very shallow local energy minimum, in contrast to the meta-stable APB energies of the classical potential [Fig. 5(b)]. Besides, the NEP reproduces the entire γ-surface with energy ordering of SISF < CSF < APB [Fig. 5(d)] as illustrated by DFT, but the widely used EAM-Zope potential has failed to capture this ordering.

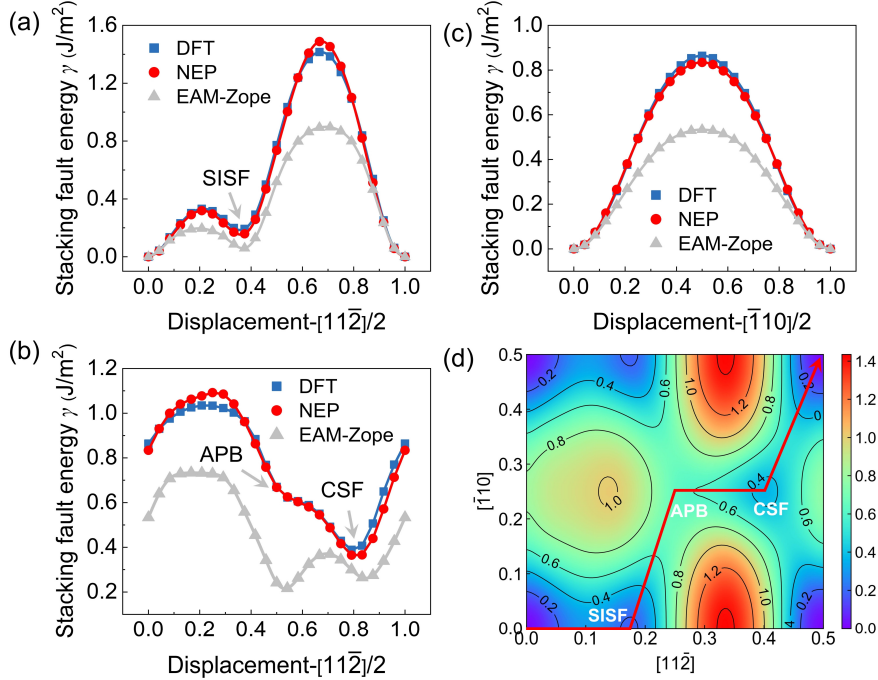

**FIG. 5.** The generalized stacking fault energy on the {111} plane of γ-TiAl predicted by NEP, EAM-Zope, and DFT. (a) The γ-line along the [11$\bar{2}$] direction passing through the SISF. (b) The γ-line along the [11$\bar{2}$] direction passing through the APB and CSF. (c) The γ-line along the [$\bar{1}$10] direction. (d) The NEP calculated γ-surface of {111} plane. The red solid arrow indicates the ideal slip path.

The α$_2$-Ti$_3$Al phase with hexagonal symmetry has multiple slip phases. Here, we mainly focus on the basic {0001} plane, and the results of stacking fault energies are shown in Fig. 6. The NEP reproduces the DFT γ-lines along [01$\bar{1}$0] [Fig. 6(a) and (b)] and [$\bar{2}$110] [Fig. 6(c)] directions. We note that the γ-lines from 0 to 40% along the [01$\bar{1}$0] direction are not well-reproduced [Fig. 6(b)], but this is not relevant since the energy is very high and is not along the minimum energy path for slip. More importantly, the NEP has accurate SISF, APB, and CSF energies as listed in Table VII; it slightly overestimates SISF by nearly 40%, while providing a more accurate prediction of APB and CSF within 3% from the DFT values. Furthermore, the unstable stacking faulting energies γ$_{us}$ along the [01$\bar{1}$0] direction are also reproduced by NEP [Fig. 6(a)]. These unstable stacking faulting energies govern the dislocation nucleation barriers on the basal plane. For a clear illustration, we have presented the accurate γ-surface predicted by NEP and labeled the expected slip path in Fig. 6(d). On the other hand, the EAM-Zope and EAM-Farkas substantially underestimate stacking fault energies by more than 55% (Table VII). The EAM-Zope potential even exhibits an almost negligible SISF energy, potentially leading to an unrealistically substantial separation of partial dislocations on the basal plane. The MEAM-KIM potential overestimates SISF by ~155% and underestimates APB by ~36%. Therefore, the developed NEP model is expected to possess high predictive accuracy for stacking faults and related properties.

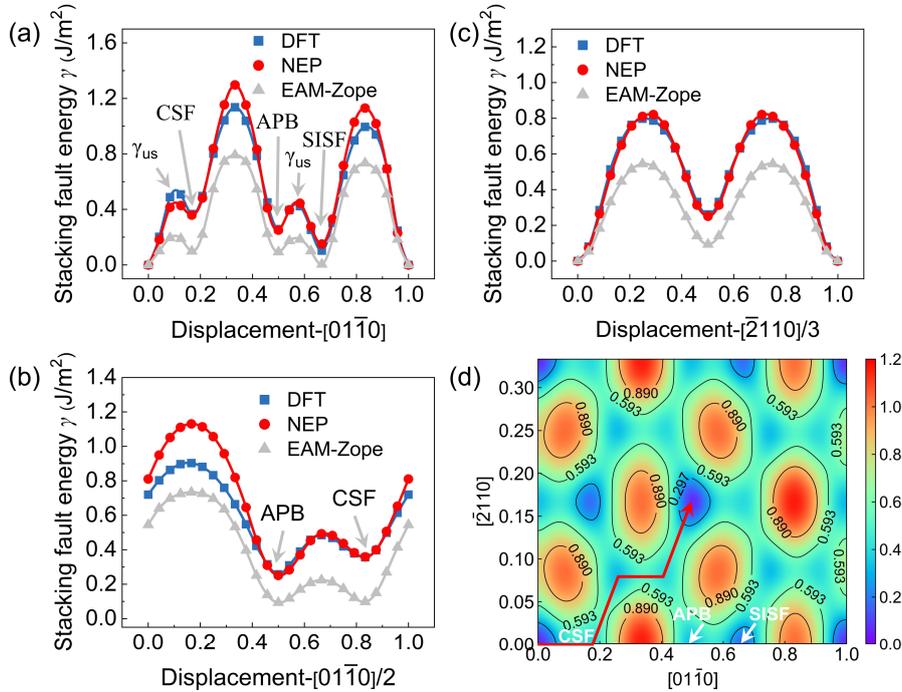

**FIG. 6.** The generalized stacking fault energy on the {0001} plane in the α$_2$-Ti$_3$Al predicted by NEP, EAM, and DFT. (a-b) The γ-line along the [01$\bar{1}$0] direction. (c) The γ-line along the [$\bar{2}$110] direction. (d) The NEP calculated γ-surface of {0001} plane. The red solid arrow indicates the ideal slip path.

**Table VII.** Stacking fault energies on the {111} plane of γ-TiAl and {0001} plane of α$_2$-Ti$_3$Al calculated by DFT, NEP, EAM and MEAM. The percentage errors with respect to DFT values are indicated in the parenthesis.

| Structure | Property | DFT | NEP | EAM-Zope | EAM-Farkas | MEAM-Kim |
|---|---|---|---|---|---|---|
| γ-TiAl | SISF (mJ/m$^2$) | 181,162[a],178[b] | 164 (9.4%) | 59 (67.4%) | 88 (51.4%) | -46 (125.4%) |
| | APB (mJ/m$^2$) | 623,635[c],640[d] | 625 (0.3%) | 216 (65.3%) | 397 (36.3%) | 149 (76.1%) |
| | CSF (mJ/m$^2$) | 388,371[e],352[f] | 364 (6.2%) | 265 (31.7%) | 343 (11.6%) | 118 (69.6%) |
| α$_2$-Ti$_3$Al | SISF (mJ/m$^2$) | 104,117[g] | 147 (41.3%) | 4 (96.2%) | 29 (72.1%) | 224 (115.4%) |
| | APB (mJ/m$^2$) | 257,256[h] | 250 (2.7%) | 93 (63.8%) | 108 (58%) | 163 (36.6%) |
| | CSF (mJ/m$^2$) | 366,320[i] | 358 (2.2%) | 98 (73.2%) | 163 (55.5%) | 324 (11.5%) |

[a]Ref. [92]; [b]Ref. [93]; [c]Ref. [94]; [d]Ref. [93]; [e]Ref. [95]; [f]Ref. [93]; [g]Ref. [96]; [h]Ref. [48]; [i]Ref . [48].

**Lattice and elastic constants at finite temperatures**

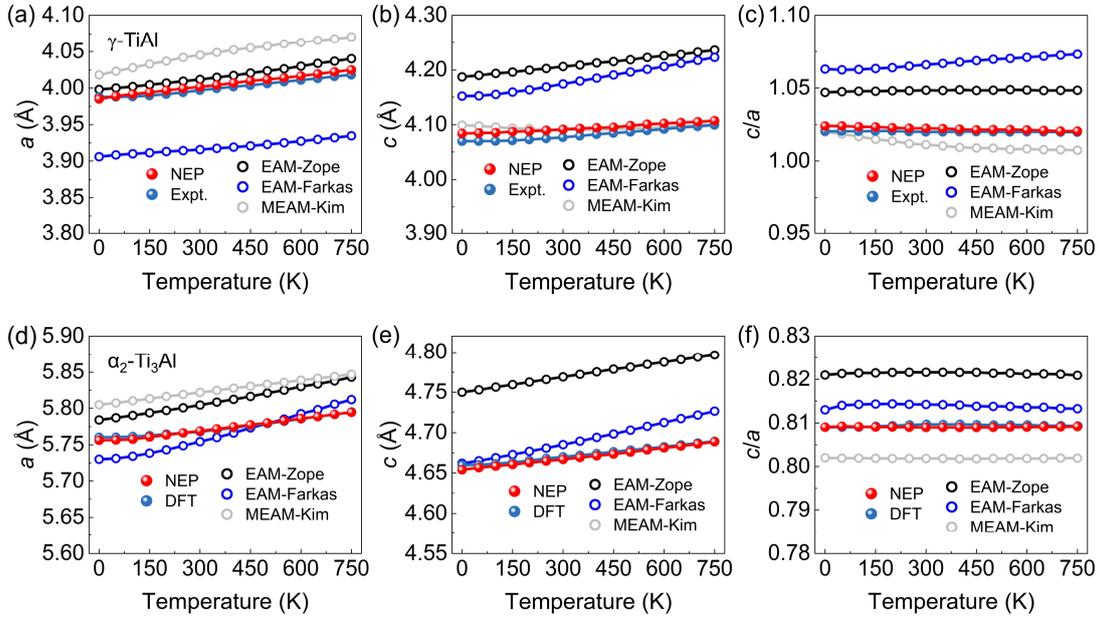

**FIG. 7.** Lattice parameters *a*, *c*, and *c/a* for γ-TiAl and α$_2$-Ti$_3$Al as a function of temperature calculated by the NEP, EAM and MEAM in comparison with available experimental [81] and DFT [97] values. (a-c) Lattice constants *a*, *c*, and *c/a* of γ-TiAl. (d-f) Lattice parameter *a*, *c*, and *c/a* of α$_2$-Ti$_3$Al.

All properties discussed above are calculated at 0 K to provide the fundamental material characteristics. Given that TiAl alloys are commonly employed in high-temperature applications across the aircraft, automotive, and power industries, it is crucial to investigate various properties of TiAl systems at finite temperatures. In typical γ-TiAl and α$_2$-Ti$_3$Al, lattice and elastic constants under finite temperatures can significantly influence material plasticity behavior. For instance, the *c/a* ratio affects dislocation behavior and slip plane spacing. Accurate elastic anisotropy properties are equally vital, impacting stress concentrations and the initiation of microcracks. Thus, we firstly study the temperature dependence of lattice parameters ($a$, $c$, $c/a$) for γ-TiAl and α$_2$-Ti$_3$Al.

As shown in Fig. 7, the lattice parameters predicted by NEP from 0 to 750 K closely match the reference experiment or DFT values for both γ-TiAl [Fig. 7(a-c)] and α$_2$-Ti$_3$Al [Fig. 7(d-f)]; the maximum difference is within 0.4% compared to reference values. The NEP model reproduces the trends observed in experiments and DFT calculations, including the temperature-dependent increase in lattice constants $a$ and $c$, as well as the nearly constant variation of $c/a$. For EAM and MEAM potentials, their overall predictions for the lattice constants of the two structures tend to be less accurate. The EAM-Zope potential provides relatively accurate lattice constant $a$ for γ-TiAl [Fig. 7(a)] and α$_2$-Ti$_3$Al [Fig. 7(d)] (within 1% error), but significantly overestimates $c$ and $c/a$ by more than 2% compared to reference values. The EAM-Farkas potential underestimates lattice constant $a$ in both phases (1-2%). Among these classical potentials, the MEAM-Kim potential has the best overall agreement with reference values with errors within 1%.

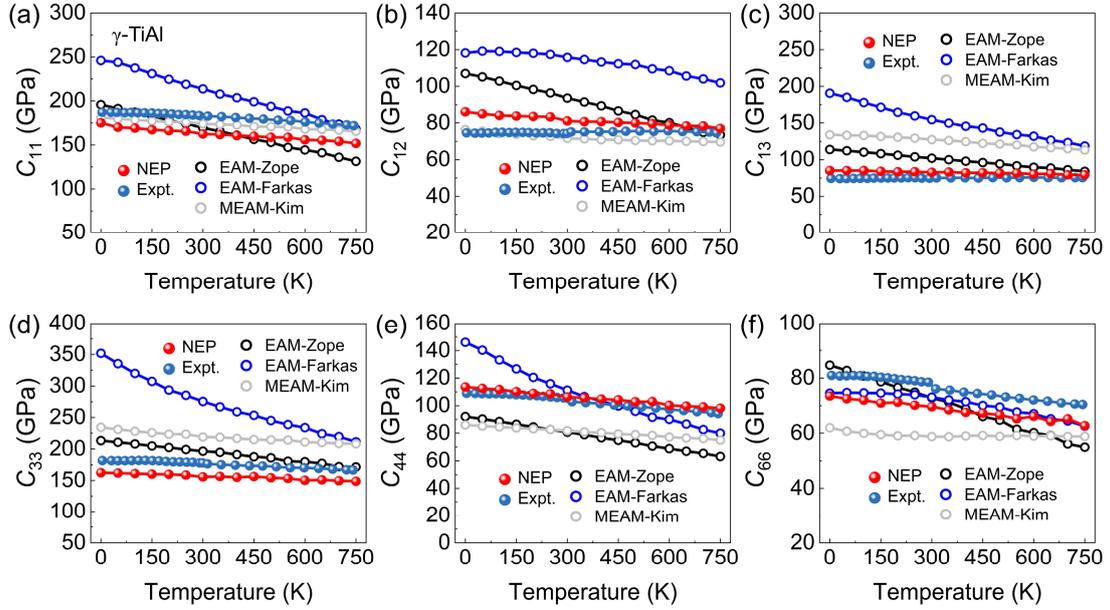

**FIG. 8.** Temperature-dependent elastic constants of γ-TiAl predicted by the NEP, EAM, and MEAM in comparison with experimental values [81,90].

Figures 8 and 9 illustrate the elastic constants of TiAl systems as functions of temperature in comparison with available experiments or DFT values. The finite temperature elastic constants $C_{ij}$ are measured with the averaging global stress method, using positive and negative strains of 2% at each temperature. The global stress is computed by averaging the stresses sampled over 6 ps (6000 time-steps) after equilibration. Results show that all interatomic potentials calculated elastic constants decrease with increasing temperature for both phases. For γ-TiAl [Fig. 8], similar to the cases at 0 K, the NEP calculated elastic constants are accurate in the entire temperature range when compared to experimental values. Minor discrepancies are observed in $C_{12}$ [Fig. 8(b)] and $C_{13}$ [Fig. 8(c)], with errors of ~15% and 14%, respectively. The NEP also accurately reproduces the elastic constants of Ti$_3$Al at different temperatures [Fig. 9], and the largest discrepancies lie in the values of $C_{44}$ with an error of 12% [Fig. 9(e)].

On the other hand, the temperature-dependent elastic constants predicted by these classical interatomic potentials cannot simultaneously accommodate both phases and exhibit considerable deviations from experimental or DFT values. The MEAM-Kim potential only exhibits favorable agreement with experimental values for $C_{11}$ [Fig. 8(a)] and $C_{12}$ [Fig. 8(b)] in the γ-phase, whereas EAM-Zope potential only shows a good description of the α$_2$ phase. In general, the classical potentials have substantial deviations in characterizing the elastic constants of the complex γ and α$_2$ phases when compared to simple FCC or BCC elemental metals, with errors surpassing 100% in specific components. The above discussions highlight the significant advantage of the NEP model in describing lattice and elastic constants under finite temperatures for complex TiAl intermetallic compounds.

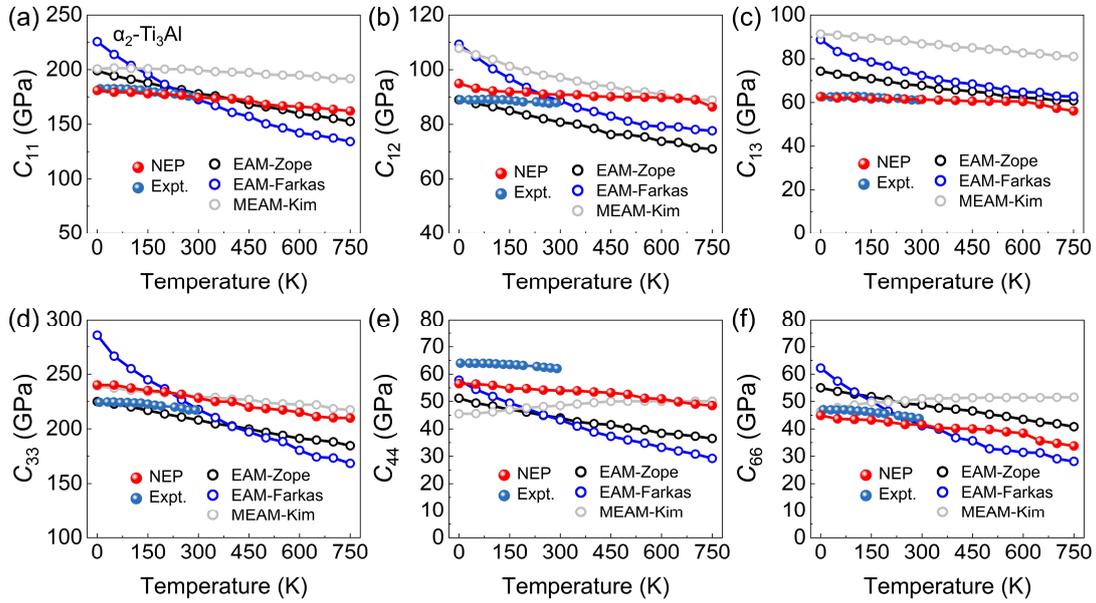

**FIG. 9.** Temperature-dependent elastic constants of α$_2$-Ti$_3$Al predicted by the NEP, EAM, and MEAM in comparison with experimental values [82].

**Melting points**

We then calculate the melting point by solid-liquid coexistence method with large-size supercells. The solid-liquid biphasic systems with half of the atoms in the liquid phase and the other half in the solid phase are used for MD simulations under zero pressure in the *NPH* ensemble. The simulation box contains a 40×20×20 supercell with ~50000 atoms for each system. If the overall system is maintained at a temperature slightly below the melting point, a fraction of the liquid phase will solidify, releasing the necessary latent heat to warm up the system. Conversely, if the system exceeds the melting temperature, the latent heat essential for solid melting will induce cooling. The solid-liquid interface assists the nucleation for the melting or solidification process. The temperature at which the initial equilibrium is reached between the solid and liquid phases without any interface motion is identified as the melting point.

**Table VIII.** The NEP calculated melting points $T_m$ for elemental and binary structures in comparison with experimental values. The percentage errors with respect to experimental values are indicated in the parenthesis.

| System | $T_m$-NEP (K) | $T_m$-Expt. (K) | Error |
|---|---|---|---|
| Ti | 1671 (± 5) | 1941 [a] | 13.9% |
| Al | 874 (± 5) | 935 [b] | 6.5% |
| Nb | 2532 (± 5) | 2742 [c] | 7.6% |
| γ-TiAl | 1776 (± 5) | 1733 [d] | 2.5% |
| $α_2$-$Ti_3Al$ | 1803 (± 5) | 1873 [d] | 3.7% |
| $D0_{22}$-$TiAl_3$ | 1662 (± 5) | 1623 [d] | 2.4% |

[a]Ref. [98]; [b]Ref. [99]; [c]Ref. [80]; [d]Ref. [100].

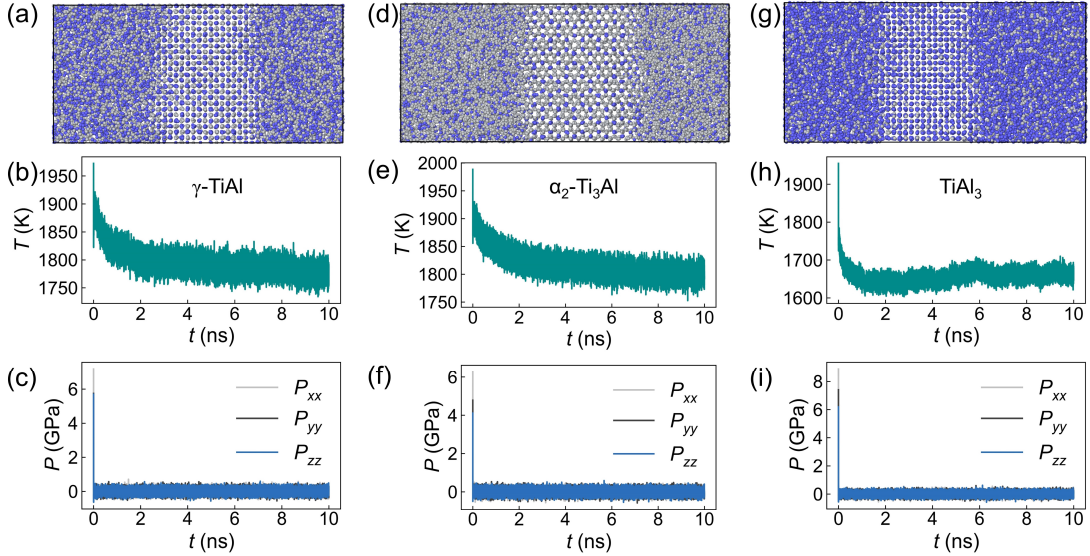

**FIG. 10.** The NEP calculated melting point with solid-liquid coexistence method of (a-c) γ-TiAl, (d-f) $α_2$-$Ti_3Al$, and (g-i) $D0_{22}$-$TiAl_3$. In each subfigure, the topmost figure represents a snapshot of the equilibrium solid-liquid interface structure; the middle figure illustrates the evolution of temperature $T$ as a function of time $t$; the bottom figure illustrates the variation of pressure $P$ ($P_{xx}$, $P_{yy}$ and $P_{zz}$) as a function of time $t$. In structure snapshots, the gray and blue atoms represent Ti and Al, respectively.

Figure 10 illustrates the evolution of temperature and pressure over time for typical Ti-Al binary systems, along with corresponding snapshots of the equilibrium solid-liquid coexistence structures. The results of melting point for elemental systems are presented in Fig. S1 in Supplemental Material. After temperature and pressure have converged, we can statistically determine the value of the melting point. As summarized in Table VIII, the NEP model accurately predicts the melting points $T_m$ for all binary Ti-Al structures, showing excellent agreement with experimental results, with a maximum error of less than 4%. For instance, the $T_m$ of γ-TiAl predicted by the NEP is 1776 ± 10 K, closely aligning with the experimental value of 1733 K. While the NEP model systematically underestimates the melting points for all elemental structures, they remain in good agreement with the experimental values. The NEP model has a large deviation of ~14% for Ti when compared to the experimental value. However, we found that the previously reported moment tensor potential also underestimates the melting point for Ti and the predicted value is ~1667 K [101], which is very close to the melting point of 1671 K predicted by the NEP model.

**Interface structures and energies**

Typical PST TiAl single crystals comprise γ-TiAl and $α_2$-Ti$_3$Al lamellae with twin boundary (γ/γ) and phase boundary (γ/$α_2$). The lamellar TiAl alloys exhibit improved mechanical properties compared to the single-phase crystals and their deformability and strength strongly depend on the microstructures of different interfaces. The sharp γ/γ and γ/$α_2$ interfaces are perpendicular to the γ-[111] and/or $α_2$-[0001] directions. Taking equivalent rotations into account, three different γ/γ interfaces can be formed by rotating half of the γ supercell around [111] direction at intervals of 60°. As illustrated in Fig. 11(a), these interfaces include the pseudo-twin (PT), obtained by 60° rotation, the rotational boundary (RB) by 120° rotation, and the true-twin (TT) by 180° rotation. In the case of γ/$α_2$ interfaces, the experimentally observed Blackburn orientation relationships are $(111)_γ$//$(0001)_{α2}$ and $<110>_γ$//$<1120>_{α2}$. Thus, six geometrically equivalent rotational variants of γ/$α_2$ interfaces can be created by rotating the $α_2$ phase around [0001] direction with the γ phase as a matrix [see Fig. 11(b)].

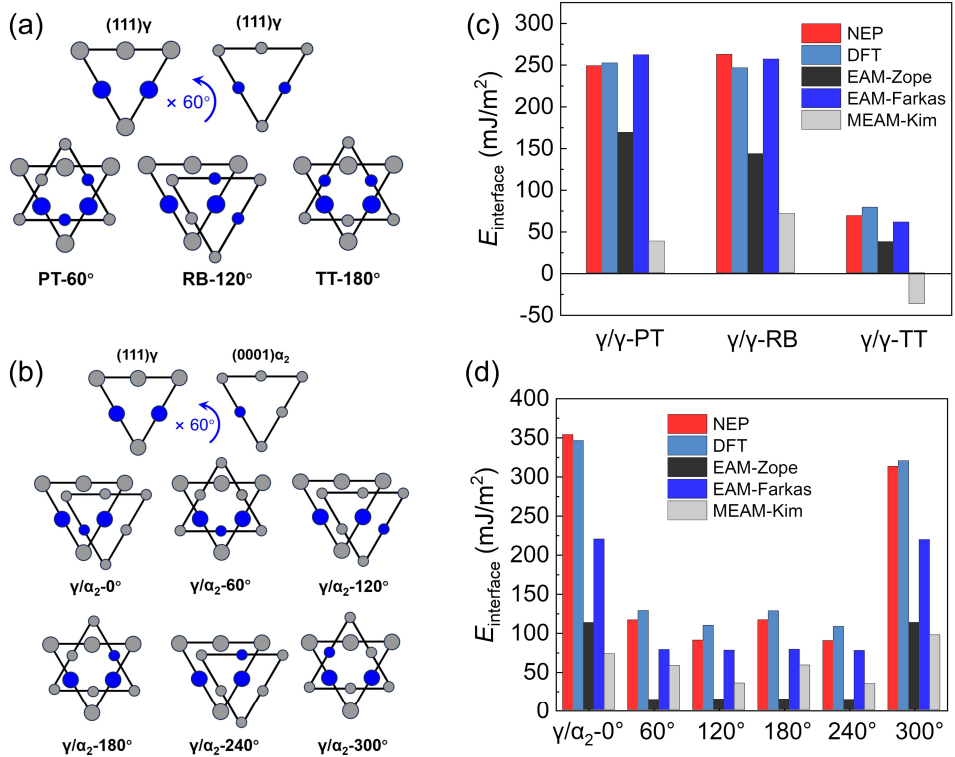

**FIG. 11.** Interface structures and energies calculated by DFT, NEP model, EAM-Zope, EAM-Farkas, and MEAM-KIM potentials. (a) Three different γ/γ interface structures obtained by rotating half of the γ phase on (111) plane around [111] direction in steps

of 60°. (b) Six different (111)γ//(0001)α₂ interface structures obtained by rotating α₂ phase around [0001] direction in steps of 60°. (c) The interface energies of γ/γ interface structures. (d) The interface energies of γ/α₂ interface structures.

We then use the general approach to calculate the interface energies of above interface structures via

$$\gamma_{GB} = \frac{E_{\text{tot}}^{\text{GB}} - E_{\text{tot}}^{\text{bulk}}}{2A}, \quad (12)$$

where $E_{\text{tot}}^{\text{GB}}$ and $E_{\text{tot}}^{\text{bulk}}$ represent the total energy of grain boundary supercell and bulk supercell, respectively. The bulk supercell contains the same number of atomic layers as the grain boundary supercell. For the γ/α₂ interfaces, the bulk energy is the summation of energies of the fully relaxed γ and α₂ phases. The results of relaxed interface energies calculated by NEP, DFT, EAM, and MEAM potentials are presented in Fig. 11(c) and (d). Among the three γ/γ variants [Fig. 11(c)], all calculations indicate that the TT interface has the lowest energy. Therefore, the TT interface is expected to be the most stable and frequently occurring boundary in lamellar microstructures, aligning with experimental observations [102,103]. The NEP predicted energies of the γ/γ interfaces show excellent agreement with the reference DFT values, with an absolute error within 20 mJ/m². Similarly, the EAM-Farkas potential also provides a good description of the energies of γ/γ interfaces, with an absolute error within 30 mJ/m². However, the interface energies calculated by EAM-Zope and MEAM-KIM potential are notably lower than the corresponding DFT values. It is noteworthy that the MEAM-KIM potential has failed to describe the energy of the TT interface, yielding an unreasonably negative value.

For the γ/α₂ interfaces, the NEP model reproduces the interface energies of these six variants in excellent agreement with the target DFT values; the difference in energy is smaller than 30 mJ/m² [Fig. 11(d)]. However, both EAM and MEAM potentials notably underestimate the γ/α₂ interface energies, with a minimum deviation of more than 30% when compared to DFT values. Among the six γ/α₂ variants, from an energy perspective, all calculations indicate that we can approximately categorize these variants into two subclasses of low-energy configurations (γ/α₂-60°, γ/α₂-120°, γ/α₂-180°, and γ/α₂-240°) and high-energy configurations (γ/α₂-0°, γ/α₂-300°). In each subclass, the included configurations can be considered as energetically degenerate variants. The γ/α₂-120° variant has the lowest energy, and thus it is the most probable interface configuration to occur in real microstructures.

**Nb doped formation energy and γ-Line**

We here investigate the influence of Nb on the mechanical properties of TiAl alloys. Firstly, the formation energy calculations are performed to evaluate the variation of the phase stability on the Nb addition. The average formation energy $E_f$ is defined as:

$$E_f = \frac{E_{tot} - N_{Ti}E_{Ti} - N_{Al}E_{Al} - N_{Nb}E_{Nb}}{N}. \tag{13}$$

Here, $E_{tot}$ is the total energy of ternary alloy. $E_{Ti}$, $E_{Al}$, and $E_{Nb}$ represent the total energies of stable elemental crystals for Ti, Al, and Nb, respectively. $N_{Ti}$, $N_{Al}$, and $N_{Nb}$ are the corresponding number of atoms, respectively. We use $Nb_{Ti}$ and $Nb_{Al}$ to denote the Nb atom occupying the Ti or Al site in γ-TiAl and α$_2$-Ti$_3$Al. As shown in Fig. 12(a), the NEP predicted $E_f$ of γ-TiAl match well with the DFT values for both cases of $Nb_{Ti}$ and $Nb_{Al}$, showing increased tendency with Nb concentration. Meanwhile, the increased $E_f$ with 18 at.% $Nb_{Ti}$ solute is only ~0.03 eV/atom, which is smaller than that of 0.10 eV/atom in $Nb_{Al}$. These results are consistent with previous simulations [18-20,104] and experimental results [105,106], indicating that increasing Nb concentration slightly decreases the phase stability of γ-TiAl, and Nb atoms are more tend to occupy the Ti site. However, the $E_f$ predicted by the EAM-Farkas potential exhibits an opposed trend. For α$_2$-Ti$_3$Al [Fig. 12(b)], the $E_f$ predicted by the NEP model also matches well with the DFT results. Additionally, both DFT and NEP show that the Nb occupying the Al site may lead to a decrease in phase stability, whereas Nb occupying the Ti site has a stabilizing effect on α$_2$-Ti$_3$Al. However, the EAM-Farkas potential fails to predict the variation tendency for $Nb_{Al}$. Herein, for both phases, Nb occupying the Ti site represents the energy-preferable situation.

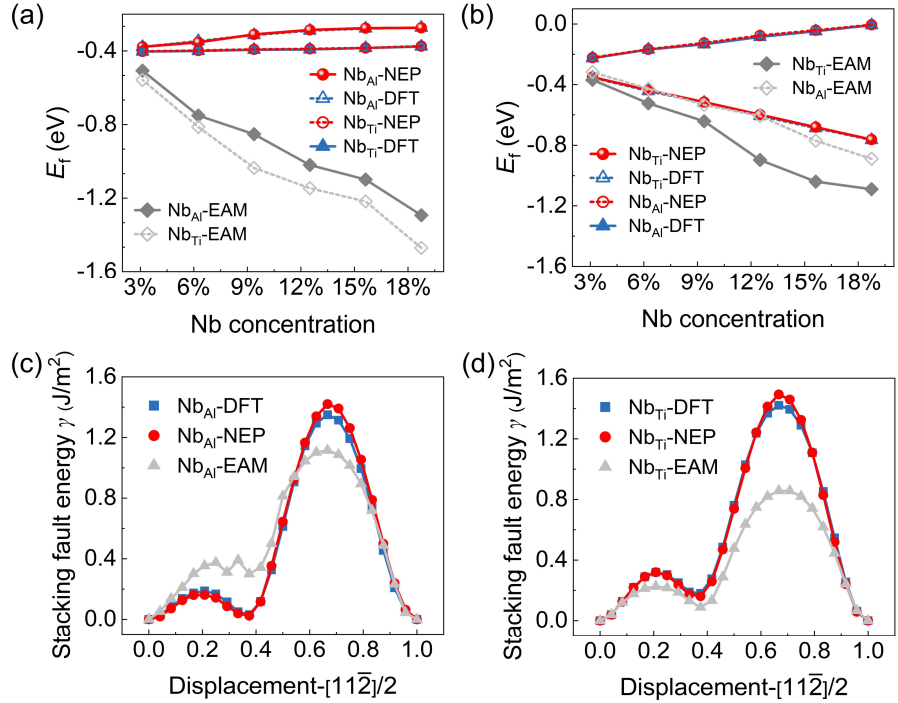

**FIG. 12.** Nb-doped formation energies and stacking fault energies. The average formation energy $E_f$ of Nb-doped (a) γ-TiAl and (b) α$_2$-Ti$_3$Al as a function of Nb concentrations calculated by DFT, NEP, and EAM-Farkas. The γ-line along the [11$\bar{2}$] direction of Nb-doped γ-TiAl (4.17 at.% Nb addition) for the case of (c) $Nb_{Al}$ and (d) $Nb_{Ti}$, respectively.

To investigate the effects of Nb on dislocation mobility, one of the most effective ways is to examine the inherent correlation of the Nb additions and the general stacking fault energy. In general, a high general stacking fault energy implies high resistance and low mobility of dislocations. The reduction (enhancement) of dislocation resistance can directly lead to the ductile (strengthening) behavior in nanomaterials. Thus, we take γ-TiAl as an example to study the substitution of Nb on the stacking fault energy of TiAl alloy. Nb atoms are randomly substituted into the Ti or Al sites of γ-TiAl supercell, with a Nb concentration of 4.17 at.%. The γ-lines along the $[11\bar{2}]$ direction are subsequently calculated using NEP, DFT, and EAM-Farkas. As shown in Fig. 12(c) and (d), the γ-lines predicted by NEP closely match the reference DFT values. The corresponding NEP SISF values for $Nb_{Al}$ [Fig. 12(c)] and $Nb_{Ti}$ [Fig. 12(d)] are 31.4 and 158 mJ/m$^2$, respectively. These results indicate that Nb occupying Al sites could significantly decrease the stable and unstable stacking fault energies when compared to γ-TiAl without Nb (SISF value is 164 mJ/m$^2$). However, the results from EAM-Farkas potential show that Nb occupying the Al sites increases the stacking fault energies, suggesting a limitation in the ability of this EAM potential to effectively capture the deformation behavior of γ-TiAl doped with Nb atoms.

**Tensile performance**

TiAl intermetallic compounds, as crucial lightweight structural materials, are frequently exposed to external tensile loading environments. Therefore, investigating their tensile mechanical properties is an essential aspect of validating the usability of the developed NEP model. We then study the uniaxial tensile properties of TiAl alloys at different temperatures. Figure S2 illustrates the stress-strain curves of γ-TiAl [Fig. S2(a)] and $α_2$-Ti$_3$Al [Fig. S2(b)] under uniaxial tensile loading along the *c*-axis direction. Initially, periodic supercell systems comprising ~64000 atoms are relaxed for 1 ns under the *NPT* ensemble at zero pressure. Subsequently, the relaxed systems are subjected to deformation with a constant rate of $4\times10^8$ s$^{-1}$. The stress-strain curves of both phases exhibit similar brittle failure behavior, featuring nearly linear elastic deformation followed by plastic yielding with a rapid drop in stress. As the temperature increases, the yielding strength for both phases gradually decreases. We obtain the temperature-dependent Young's moduli by fitting linear elastic strain region (≤0.02) [Fig. S2(c) in Supplemental Material]. The Young's moduli of γ-TiAl and $α_2$-Ti$_3$Al at 300 K are approximately 104 and 190 GPa, respectively. Notably, these values of Young's moduli demonstrate minimal sensitivity to temperature variations, consistent with experimental observations.

Despite single-phase TiAl alloys having poor room temperature ductility, the PST TiAl single crystal with controlled lamellar orientations has superior performance in both strength and ductility by introducing twin boundaries (TBs) and phase boundaries (PBs) [7]. We then choose PST TiAl single crystal as model materials to further study the mechanical properties of lamellar TiAl structures with the NEP model. The lamellar TiAl single crystal consists of a majority of the γ phase and a minority of the $α_2$ phase. The Blackburn orientation relationships based on the experimental observations are

adopted to build the lamellar model and can be described as:

$$x = [1\bar{2}10]_{\alpha_2} \parallel [1\bar{1}0]_{\gamma},\ y = [\bar{1}010]_{\alpha_2} \parallel [\bar{1}\bar{1}2]_{\gamma},\ z = [0001]_{\alpha_2} \parallel [111]_{\gamma}. \tag{14}$$

Here, we consider two simulation cases with a same TBs/PBs ratios of 1:1 but with different Nb concentrations. Case I represents the pure lamellar TiAl model without Nb; case II represents the lamellar model with 8 at.% Nb doping. Simulation cases I and II both contain 1075008 atoms, with a same supercell size of 30.3×58.3×10.3 nm³. The tensile loading is applied at 300 K along the $y$ direction with a strain rate of $4\times10^8$ s$^{-1}$. As shown in Fig. 13(a), the stress-strain curves of both lamellar models contain multiple stress drops followed by stress increase, contrasting with the behavior observed in single-phase TiAl alloys. Once the stress of the lamellar model surpasses the critical stress, the stress-strain curve will enter the flow stress stage, which is consistent with previous MD results [107]. The multiple stress-increasing stages observed in the flow stress stage may be caused by complex reasons, including interface obstruction [108-110], dislocation junctions [111,112] and dislocation retractions.

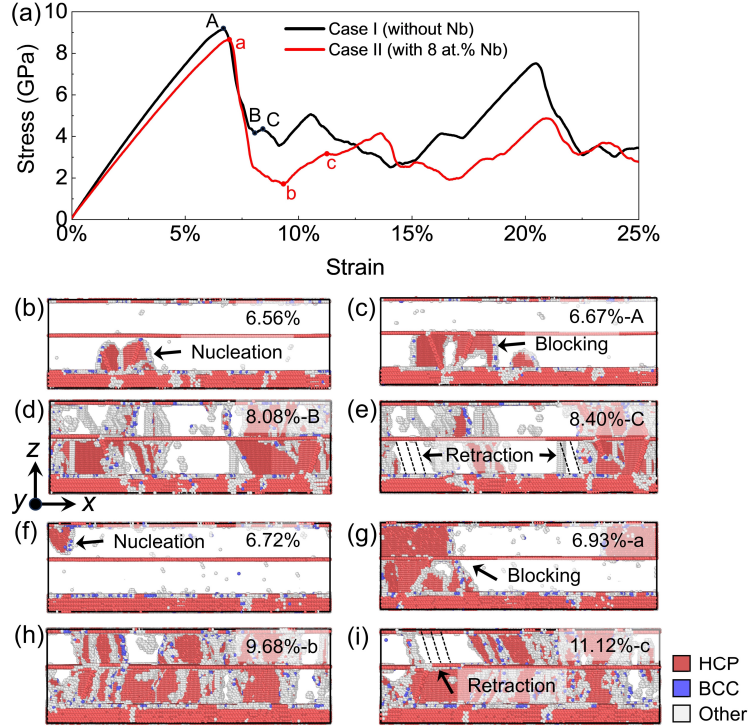

**FIG. 13.** The tensile performance of lamellar TiAl models. (a) NEP calculated stress-strain curves of two different lamellar models under uniaxial tension along the $y$ direction at 300 K. (b-e) Structure snapshots of the lamellar TiAl model without Nb at strains of (b) 6.56%, (c) 6.67%, (d) 8.08%, (e) 8.40%. The strains of 6.67%, 8.08%, and 8.40% correspond to points A, B, and C labeled on the stress-strain curve of case I, respectively. (f-i) Structure snapshots of the lamellar TiAl model with 8 at.% Nb at strains of (f) 6.72%, (g) 6.93%, (h) 9.68%, (i) 11.12%. The strains of 6.93%, 9.68%, and 11.12% correspond to points a, b, and c labeled on the stress-strain curve of case II, respectively. The defect structures are clarified by the common neighbor analysis in OVITO [113]. The flawless configurations in γ-TiAl are deleted for clarification.

The underlying mechanism of multiple stress increases can be rationalized by looking into the microstructure evolutions of lamellar models. For simulation case I, the γ-TiAl lamellae undergo plastic deformation at a strain of 6.56 % through nucleating dislocation and stacking fault from the PBs [Fig. 13(b)]. As the strain reaches 6.67% at point A, more dislocations nucleation occurs, and some dislocations are hindered by TBs [Fig. 13(c)], where the stress reaches its maximum value. Increasing the strain further leads to the dislocation transfer of the TBs in the γ phase [Fig. 13(d)], which corresponds to the decreasing of stress from points A to B. When the strain reaches 8.40% at point C, dislocation retractions occur again in γ [Fig. 13(e)], which corresponds to the stress-increasing stage from points B and C. In comparison to simulation case I without Nb, the addition of 8 at.% Nb increases the yield strain from 6.67 to 6.93% and decreases the yield stress from 9.15 to 8.64 GPa [Fig. 13(a)]. Structure snapshots under different strains show the similar behavior of defect nucleation and defect-boundary interactions [Fig. 13(f-i)]. Once surpassing the elastic limit, dislocations nucleate from the PBs and leave dislocation trails along the PBs. As Shockley partial dislocations approach TBs, they are hindered by the TBs, leading to the formation of new Shockley partial dislocations on the TBs. Throughout the entire strain process, both the nucleation and retraction of dislocations occur.

**Computational cost of NEP**

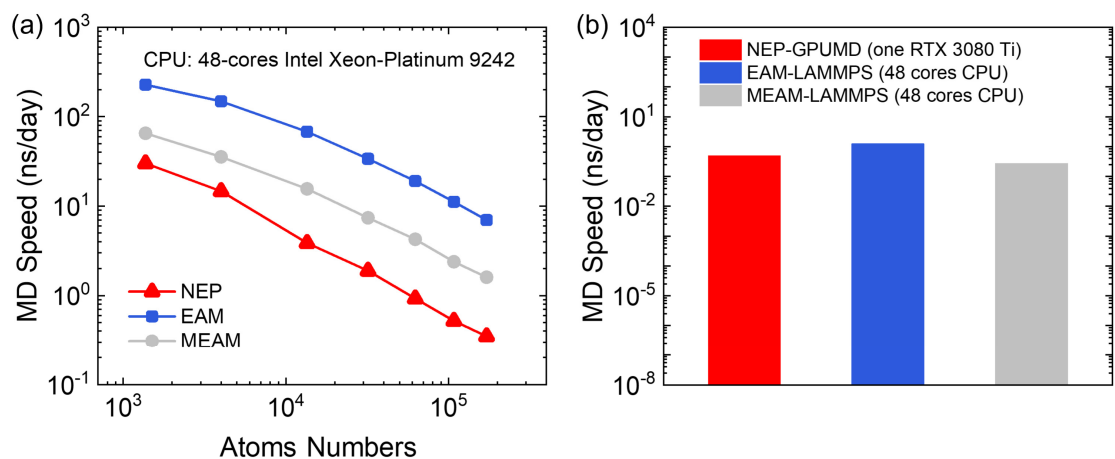

**FIG. 14.** MD Speed comparison of the NEP model, EAM-Zope and MEAM-KIM potential on (a) CPU and (b) GPU. All simulations in subplot (b) contain one million atoms. Note that MEAM potential is currently not ported to GPU in LAMMPS. Thus, we compute the MD speed for EAM and MEAM both on CPU.

The computational performance is crucial for MLPs when performing large-scale MD simulations, particularly for alloy systems that typically require modeling the mechanical properties with several million atoms. We herein compare the computational speed of NEP, EAM, and MEAM implementations on both CPUs and GPUs [Fig. 14]. On CPUs with the LAMMPS package, the NEP model is 10-20 times slower than the EAM-Zope potential and only 2-4 times slower than the MEAM-KIM potential [Fig. 14(a)]. Meanwhile, all potentials show a linear scaling with the number of atoms. On GPUs, the NEP model as implemented in GPUMD package with one single Nvidia GeForce RTX 3080 Ti GPU can reach the model size of 1 million atoms with a computational speed of 0.48 ns/day ($5.3 \times 10^6$ atom step/s), which is only two times lower than that of EAM potential (1.22 ns/day) and faster than MEAM potential (0.27 ns/day) as implemented in LAMMPS using 48 CPUs [Fig. 14(b)]. Running on 4 RTX 3080 Ti GPUs, the NEP model can reach a simulation size of 10 million atoms with a computational speed of $1.5 \times 10^7$ atom step/s. The multiple GPUs parallel efficiency of the NEP model as implemented in GPUMD can reach up to ~80%, which is significantly higher than that of ~50% for EAM potential as implemented in the LAMMPS GPU version. By synergizing high computational speed with superior parallel efficiency, the NEP model excels in performing large-scale MD simulations of TiAl systems, such as dislocation dynamics and the phase transition processes. The NEP model offers an excellent balance between accuracy and speed, surpassing the capabilities of DFT for simulations on scales previously deemed unattainable.

## IV. CONCLUSIONS

We have presented a procedure for constructing Ti-Al-Nb ternary machine-learned potential by integrating the NEP framework with an active learning sampling scheme. By fitting to an extensive dataset of bulk and defect structures, the trained potential demonstrates *ab initio* accuracy in predicting energy, force, and virial. The NEP model not only accurately reproduces the basic lattice and defect properties of elemental, binary, and ternary systems but also effectively describes material behavior at finite temperatures, encompassing aspects such as lattice thermal expansion, elastic constants, and melting points. For two typical TiAl intermetallic compounds, our NEP model successfully reproduces the generalized stacking fault energy curves and surfaces on basal planes and accurately captures the interface energy characteristics between different phases, laying a solid foundation for studying the plastic deformation of TiAl systems. Furthermore, the NEP model can also accurately reproduce the effects of Nb doping on the formation energies and stacking fault energies of TiAl alloys, facilitating deeper exploration into the influence of Nb doping on their mechanical properties. The large-scale MD simulations are also performed to explore the uniaxial mechanical properties of both single phases and lamellar structures, further validating the practical utility of the NEP model. Compared to classical EAM and MEAM potentials, the NEP model exhibits significant enhancements in accuracy when describing physical properties. More importantly, the NEP model achieves a balance between accuracy and computational speed, enabling efficient simulations of tens of millions of atoms using only a small number of GPUs. Overall, we have developed an efficient and accurate machine-learned potential for the Ti-Al-Nb ternary system, with a particular focus on high-temperature mechanical properties. The proposed training iteration procedure outlined in this work is general and applicable to train accurate interatomic potentials for a wide range of material systems. Meanwhile, this work also contributes a comprehensive dataset for benchmarking, developing, and fine-tuning interatomic potentials in ternary Ti-Al-Nb systems and higher-order TiAl-based intermetallic compound systems in the future.

## SUPPLEMENTAL MATERIAL

See Supplemental Material at http://link.aps.org/supplemental/x.xxx/PhysRevB.x.xxx.

## DATA AVAILABILITY

To ensure the reproducibility and use of the models developed in this paper, all data used in model development as well as the final fitted NEP model will be published in an open repository (https://github.com/Ricky-Zhao/NEP-TiAlNb-Potential).

## ACKNOWLEDGEMENTS

This work was supported by National Key Research and Development Program of China (2019YFA0705400), National Natural Science Foundation of China (1221101035, 12225205, 22073048, T2293691), and a Project by the Priority Academic Program Development of Jiangsu Higher Education Institutions. The computations were in part performed at the High-Performance Computational Center at NUAA. The authors would like to thank Dr. Rui Zhao at Hunan University, Dr. Yu Li at Shanghai University and Dr. Yiqi Zhu at Nanjing University of Aeronautics and Astronautics for their valuable discussion.